\documentclass[%
a4paper,       
UKenglish,     
DIV10,          
useAMS, referee, doublespace]%
{scrartcl}\usepackage[]{graphicx}\usepackage[]{xcolor}
\makeatletter
\def\maxwidth{ %
  \ifdim\Gin@nat@width>\linewidth
    \linewidth
  \else
    \Gin@nat@width
  \fi
}
\makeatother

\definecolor{fgcolor}{rgb}{0.345, 0.345, 0.345}

\usepackage{framed}
\makeatletter
 {\par\unskip\endMakeFramed%
 \at@end@of@kframe}
\makeatother

\definecolor{shadecolor}{rgb}{.97, .97, .97}
\definecolor{messagecolor}{rgb}{0, 0, 0}
\definecolor{warningcolor}{rgb}{1, 0, 1}
\definecolor{errorcolor}{rgb}{1, 0, 0}
\newenvironment{knitrout}{}{} 

\usepackage{alltt}

\usepackage[a4paper, total={6in, 9.25in}]{geometry}
\newcommand{\hlrev}{\textcolor{black}}

\usepackage{xr-hyper}
\usepackage{authblk}    
\usepackage{hyperref}
\input{header.sty}
\usepackage{color, colortbl}
\usepackage{mathtools}

\usepackage{chngcntr}
\usepackage{tikz}
\usetikzlibrary{arrows.meta, shapes, arrows}

\tikzstyle{block} = [rectangle, draw,
    text width=3em, text centered, rounded corners, node distance=3cm]
\tikzstyle{block2} = [rectangle, draw,
    text width=11em, text centered, rounded corners, node distance=7cm]
  
\tikzstyle{plain2} = [draw = none, fill = none,
text width=5.5em, minimum height = 1cm]
\tikzstyle{cloud} = [draw, ellipse, node distance=5cm,minimum height=5em]
\tikzstyle{line} = [draw, -latex]
\tikzstyle{plain} = [draw=none, fill=none]

\newcolumntype{C}[1]{>{\centering\arraybackslash}p{#1}}
\IfFileExists{upquote.sty}{\usepackage{upquote}}{}
\begin{document}

\title{\hlrev{Assessing replicability with the sceptical $p$-value:} 
Type-I error control and sample size planning} 
\ifcase\blinded 
\author{} \or 
\author{Charlotte Micheloud$^{1}$\footnote{Corresponding author,  \texttt{charlotte.micheloud@uzh.ch}}, Fadoua Balabdaoui$^2$ and Leonhard Held$^1$}
\affil{\large $^1$University of Zurich\\
Epidemiology, Biostatistics and Prevention Institute (EBPI)\\
  and Center for Reproducible Science (CRS) \\
  Hirschengraben 84, 8001 Zurich, Switzerland\\ 
  and \\
$^2$ETH Zurich\\
Seminar für Statistik \\
Rämistrasse 101, 8092 Zürich, Switzerland}
\date{\large \today}
\fi

\maketitle
\vspace{-.5cm}
\begin{center}
\begin{minipage}{12cm}
  \textbf{Abstract}: We study a statistical framework
  for replicability based on a recently proposed
 quantitative measure of replication success, the sceptical $p$-value. 
 A recalibration is proposed to obtain
  exact overall Type-I error control if the effect is null in both studies
  and additional bounds
  on the partial and conditional Type-I error rate, which represent the case 
  where only one study has a null effect. The approach avoids the
  double dichotomization for significance of the two-trials rule and
  has larger project power to detect existing effects 
  over both studies in combination. 
  It can also be used for power calculations
  and requires a smaller replication sample size than the two-trials rule
  for already 
  convincing original studies.
We illustrate the performance of the proposed methodology
  in an application to data from the Experimental Economics Replication
  Project.  
\end{minipage}
\end{center}


\doublespacing


\section{Introduction}\label{sec:introduction}

Replication plays a key role to build confidence in the scientific
merit of published results.
The so-called replication crisis has
led to increased interest in replication studies over the last decade
\citep{knaw2018,NAS2019} with
large-scale replication projects being conducted in various fields
\citep{OSC2015,Camerer2016,Camerer2018,Errington2021}.  Deciding
whether a replication is successful is, however, not a straightforward
task, and different statistical methods are currently being used. For
example, the Reproducibility Project: Cancer Biology
\citep{Errington2021}, an 8-year effort to replicate experiments from
high-impact cancer biology papers, has used no less than seven different methods
to assess replicability, including significance of both the original and
replication study, compatibility of the original and replication
effect estimates, and computation of a
meta-analytic combined effect estimate with confidence interval.

Declaring a replication as successful if both the original and
replication study are significant at level $\alpha$ (usually one-sided
0.025) is known as the two-trials rule in drug development and
serves as a useful benchmark.
Specifically, it has an overall Type-I error (T1E) rate of $\alpha^2$
\citep{senn:2007} if the effect is null in both studies
and \hlrev{the partial T1E rate,} the risk of a false claim of replication success
if a true effect is present in at most one of the two studies\hlrev{,} is
bounded by $\alpha$ \citep{Heller_etal2014}.  However, the
  ``double dichotomization'' of the two-trials rule has serious
  limitations.
  Firstly, it is common practice to replicate interesting
  findings even if the original study does not pass the much-critized
  ``bright-line'' threshold of $\alpha = 0.025$. For example, in the
  \citet{OSC2015} Psychology replication project, four studies have
  been included despite ``falling a bit short of the [two-sided] 0.05
  criterion -- $P$ = 0.0508, 0.0514, 0.0516, and 0.0567 -- but all of
  these were interpreted as positive effects''.
  Similarly, the
  Experimental Economics Replication Project \citep{Camerer2016}
  has chosen to replicate 18 studies, two of which have not been significant
  at the conventional two-sided 0.05 standard.
Strict application of the two-trials rule, however, would make it
  pointless to try to replicate such non-significant original studies.
  Secondly, the two-trials
  rule has been shown to have {relatively low} \hlrev{project power, \ie}
  power to detect existing effects
  over both studies in combination \citep{Maca2002,held2020b}. 
  These issues suggest investigating alternative methods to assess
  replication success.


A recent proposal by \cite{held2020} combines a
reverse-Bayes approach \citep[see][for a recent review]{held_etal2022}
with a prior-predictive check for conflict \citep{box:1980} and gives rise to a
quantitative measure of replication success, the sceptical $p$-value.
The sceptical $p$-value depends on the two study-specific $p$-values,
but also on the ratio of the variances of the
original and replication effect estimates. 
The method treats
the original and replication study not as exchangeable and specifically 
penalizes
shrinkage of the replication effect estimate, compared to the original
one. 
The effect size perspective has been further explored 
to propose a modification
based on the golden ratio \citep{held_etal2020}, in the following called the 
golden sceptical $p$-value.
While significance of both studies is a necessary but not sufficient success criterion in the original formulation,
the golden sceptical $p$-value also 
allows original studies with a ``trend to significance'' to be successful at
replication, but only if the effect estimate at replication is larger
than at original. 

The golden sceptical $p$-value addresses some of the
 problems of the two-trials rule.  It can flag
  replication success if the original or replication $p$-value does
  not meet the significance threshold $\alpha$ and provides larger
  project power
  than the two-trials rule.  However, the probability for replication success
  if the observed original effect estimate is the true
  effect 
  while being non-significant is always smaller than 50\%. This
  is less extreme than the two-trials rule where non-significant
  original studies can never lead to replication success, but
  precludes sample size planning for replication studies of
  non-significant original findings at commonly used power values such
  as 80\% or 90\%.  Furthermore, neither the original (nominal) nor
  the golden sceptical $p$-value has an exact overall T1E rate of $\alpha^2$ if the effect is null in both studies. The T1E rate
  of the nominal one is always below $\alpha^2$, whereas the T1E rate
  of the golden one can exceed $\alpha^2$ if the variance ratio
  (original to replication) is smaller than one. An alternative reverse-Bayes approach based
  on Bayes factors has also been proposed, but the resulting sceptical Bayes factor 
  can also not be used for sample size planning if the original
  result was not convincing on its own \citep[Section
  3.3]{PawelHeld2022}.

In this paper we study the sceptical $p$-value from a frequentist
perspective and examine its T1E rate in greater detail. 
\hlrev{We aim to control the overall T1E rate rather than the partial 
T1E rate to allow for a fair comparison with the two-trials rule \citep{Rosenkranz2023}.
Any other method with partial T1E rate $< \alpha$ (such as the nominal 
sceptical $p$-value) will have a smaller 
overall T1E rate than the two-trials rule, 
and its success region will be a subset of the success region of the 
two-trials rule.
The ultimate goal is \hlrev{hence} to recalibrate the sceptical $p$-value to achieve exact 
overall T1E control \hlrev{at level $\alpha^2$}
and to enable sample size calculations \hlrev{also} for non-significant 
original studies.
However, any method with the same overall T1E rate as the two-trials rule
will have an increased partial T1E rate. 
This is also the case for the sceptical $p$-value, but we will 
show that the increase in 
conditional T1E rate, the risk of a false claim of success if the replication 
study is properly powered based on the original result,
is always below $2\alpha = 5$\%, which is considered 
a `sensible option' by \citet{Rosenkrantz2002}.
} 

In Section~\ref{sec:framework} we
describe the underlying statistical framework for replicability and consider 
T1E rates under two different null hypotheses, the intersection and the union
null \citep{Heller_etal2014} in Section~\ref{sec:null_hyp}.  The two-trials rule
  (Section~\ref{sec:2TR}) and the harmonic mean $\chi^2$-test
  (Section~\ref{sec:harmonic}) are identified as special cases of
  this framework with
  exact overall T1E control of $\alpha^2$ under the
    intersection null.  The relevant null distribution is then derived in all other cases and the sceptical $p$-value 
    is recalibrated in
  Section~\ref{sec:exactT1Econtrol} to achieve
  exact overall T1E control for every possible value of the variance ratio.
  Limiting cases and further properties are
  described in Sections~\ref{sec:limitingcases}
  and~\ref{sec:comparison}.  In Section~\ref{sec:rs}, the
  sceptical $p$-value is used as a dichotomous criterion for
  replication success 
 with focus on the partial T1E rate under the union null hypothesis in Section~\ref{sec:adaptiveLevel}.
The sceptical
  $p$-value and the two-trials rule are then compared in terms of
  success regions (Section~\ref{sec:combTest}), project power
  (Section~\ref{sec:pp}) and for the design of replication studies, \hlrev{with 
  particular focus on the conditional T1E rate} (Section~\ref{sec:design}).
  An application to data from the
  Experimental Economics Replication Project is given in Section
  \ref{sec:appEE}.  We close with some discussion in Section
  \ref{sec:discussion}.

\section{A statistical framework for replicability}\label{sec:framework}

Let $\hat \theta_i$ denote the estimate of the unknown effect size
$\theta_i$ and $\sigma_i$ the corresponding standard error from the
original and replication study, $i \in \{o, r\}$. 
As in standard meta-analysis we assume that the $\hat \theta_i$'s are
independent and follow a normal distribution with mean $\theta_i$ and
known variance $\sigma_i^2$.  Let
$z_i = {\hat \theta_i}/{\sigma_i}$
denote the test statistic for the null hypothesis $H^{\,i}_{0}$: $\theta_i=0$, 
$i \in {o, r}$, and $p_i = 1-\Phi(z_i)$ the 
corresponding one-sided $p$-value for the alternative $H^{\,i}_{1}$: 
$\theta_i>0$, 
here $\Phi(.)$ denotes the standard 
normal cumulative distribution function. 
Replication success \hlcm{at level $\gamma$} is achieved if 
\begin{equation}\label{eq:general}
\left({z_o^2}/{z_{\gamma}^2}-1\right)_{+} 
  \left({z_r^2}/{z_{\gamma}^2}-1\right)_{+} \geq c 
\end{equation}
holds, here $x_{+} = \max\{0, x\}$, $c = \sigma_o^2/\sigma_r^2> 0$
is the variance ratio and
$z_\gamma = \Phi^{-1}(1 - \gamma) >0$ is the threshold \hlcm{at replication 
success level $\gamma$}.  

The two-sided
formulation only requires \eqref{eq:general},
irrespectively of the signs of the estimates $\hat \theta_o$ and
$\hat \theta_r$, but suffers from the ``replication paradox'' \citep{Ly_etal2019}
because replication success can occur even if the effect
estimates $\hat \theta_o$ and $\hat \theta_r$ are in opposite directions.
The one-sided formulation avoids this problem with
the additional requirement that the two estimates are both in the
same pre-specified (w.l.o.g.~positive) direction,
\begin{equation}\label{eq:direction}
\hat \theta_o > 0 \mbox{ and } \hat \theta_r > 0, 
\end{equation}
and so we usually require both \eqref{eq:general} and \eqref{eq:direction} 
to achieve replication success (if not stated otherwise).

The success conditions \eqref{eq:general} and \eqref{eq:direction} can
be motivated from a recent proposal to define replication success with
a two-step procedure \citep{held2020}: First, a significant original
study at \hlcm{one-sided} level $\gamma$ is challenged 
by a normal prior with mean zero
modelling the belief of a hypothetical sceptic who regards the absence
of an effect to be the most likely reality \citep{matthews2018}. The
prior variance is chosen such that 
the posterior probability
that the effect is negative is $\gamma$.  Secondly, the conflict
between the replication study result and the sceptical prior is
quantified with a prior-predictive tail probability
$p_{\mbox{\scriptsize Box}}$ \citep{box:1980}. Replication success at
level $\gamma$ is {then} achieved if
$p_{\mbox{\scriptsize Box}} \leq \gamma$, \ie if there is more conﬂict
between the sceptical prior and the replication study than there was
evidence against the null hypothesis based on the original data.

We are often interested in the smallest
possible value of $z_{\gamma}^2$ which solves
\eqref{eq:general} and denote
this value as $z_{S}^2 \in (0, \min\{z_o^2, z_r^2\})$, defined as the smallest
positive root of
\begin{equation}\label{eq:equation}
\left({z_o^2}/{z_{S}^2}-1 \right) \left({z_r^2}/{z_{S}^2} 
- 1 \right) = c. 
\end{equation}
This is a quadratic equation in $z_S^2$ and can be solved analytically.
Any ${z_S} = + \sqrt{z_S^2} \geq z_\gamma$ will hence lead to replication success
\hlcm{at level $\gamma$},
so the threshold $z_\gamma$ in \eqref{eq:general} serves as a critical value
for the test statistic $z_S$.  If the effect
estimates fulfill \eqref{eq:direction}, the transformation
$p_S = 1-\Phi({z_S})$ \soutr{now} defines the (one-sided)
sceptical $p$-value in its original formulation and the criterion ${z_S} \geq z_\gamma$
 for replication success translates to $p_S \leq \gamma$.
If \eqref{eq:direction} doesn't hold
we set $p_S=\Phi({z_S})$ \citep[Section~3.3]{held2020}.


\subsection{Null hypotheses and Type-I error rates}\label{sec:null_hyp}

The T1E rate is the probability of a false claim of
replication success under a given null hypothesis. In the replication 
setting with two studies, this probability can be considered under two
different null hypotheses \citep{Heller_etal2014}.
The \emph{intersection null hypothesis} is
a point null hypothesis, defined
as the intersection of the study-specific null hypotheses $H^{\,i}_{0}$:
$\theta_i=0$, $i=o, r$:
\begin{equation}\label{eq:H0i}
 H^{\,o}_{0} \cap H^{\,r}_{0}.
\end{equation}
The probability of a false claim of
replication success with respect to the intersection null
\eqref{eq:H0i} is the \emph{overall} T1E rate. 

The \emph{no-replicability} or \emph{union null hypothesis}
is defined as the complement of the
\mbox{alternative} that the effect is non-null in both studies. 
This is a
composite null hypothesis, which also includes the possibility that
only one study has a null eﬀect:
\begin{equation}\label{eq:H0c}
  H^{\,o}_{0} \cup H^{\,r}_{0}.
\end{equation}
The probability of a false claim of
replication success with respect to the union null \eqref{eq:H0c}, the 
\emph{partial} T1E rate, depends on the values of 
$\theta_o$ and $\theta_r$. One of them is zero but the other one may
not be zero. The partial T1E rate has been recently
  investigated by \citet{Zhan2023} for the two-trials rule.  
 In Section \ref{sec:design} we also study the T1E rate
  under $H_0^{\, r}$ only, conditional on the result of the original study. 
  This \emph{conditional} T1E rate is of \hlrev{primary} interest 
  \hlrev{because in practice} the design 
  of the replication study depends on the result from the original study
  \citep{Anderson2022}.

A necessary but not sufficient condition for the replication success 
criterion~\eqref{eq:general} to
hold is $\min\{\abs{z_o}, \abs{z_r}\} >
z_\gamma$,
as otherwise the left-hand side of \eqref{eq:general} is zero.
\hl{Combined with \eqref{eq:direction} this translates to the necessary but not sufficient
  requirement $p_{\max}  = \max\{p_o, p_r\} < \gamma$. Under the union null hypothesis,
  either $p_o$ or $p_r$ is uniform distributed,   so $\gamma$ is a bound on the
partial T1E rate}\soutr{probability of a false claim of replication success}
of the 
sceptical $p$-value for any value of
the variance ratio $c$. 
Likewise, the overall T1E rate is smaller than $\gamma^2$ due to
independence of the two studies.

This raises the question what value for $\gamma$ to use in~\eqref{eq:general}.
\hlcm{The \emph{nominal} success level is the standard significance 
level $\gamma = \alpha$,}
so controls the overall and partial T1E rate at $\alpha^2$ respectively 
$\alpha$ for every value of $c$. However, T1E control is not exact and 
the overall T1E rate can be
considerably smaller than $\alpha^2$ \citep[Section 3.2]{held_etal2020}.
\hlrev{Replication success at the nominal level is hence only 
possible if both $p$-values are smaller than $\alpha$ and is much more 
stringent than the two-trials rule.}
The \emph{golden} \hlcm{success level is}
$\gamma(\alpha)= 1-\Phi({z_\alpha}/\sqrt{\varphi}) > \alpha$, 
\hlrev{where $z_\alpha = \Phi^{-1}(1 - \alpha)$} and 
$\varphi=(\sqrt{5}+1)/2$ is the golden ratio.
\hlrev{For example, 
$\gamma(\alpha = 0.025) = 0.062$.}
 It is therefore less restrictive than the nominal \hlcm{level} 
in the assessment of replication success.
By construction,  \hlrev{it controls the partial T1E rate
at level $\gamma(\alpha)$ and} the overall T1E rate at $\gamma(\alpha)^2$ and even at
$\alpha^2$ if  $c \geq 1$  and $\alpha \leq 0.058$ 
\citep[Section 3.2]{held_etal2020}, but the actual overall T1E rate can be much 
smaller than the corresponding bound. 
\hlcm{Comparing $p_S$ 
to the golden level $\gamma(\alpha)$ is equivalent to comparing the 
 golden sceptical $p$-value $\tilde p_S = 1-\Phi(\sqrt{\varphi}\,  z_S)$
to $\alpha$.}

In what follows we describe how to obtain exact overall T1E control of the 
sceptical $p$-value for any particular value of $c$.
  This is motivated through the identification of the two-trials rule as a 
  special case of the general formulation~\eqref{eq:general} respectively~\eqref{eq:equation} and leads to a 
  \hlcm{ \emph{controlled} success level} $\gamma_c(\alpha)$
  that depends on both $\alpha$ and $c$.
\subsection{The two-trials rule}\label{sec:2TR}

The two-trials rule
requires significance of both studies at the one-sided significance level $\alpha$,
so corresponds to $z_o \geq z_\alpha$ and
$z_r \geq z_\alpha$, and translates to the single criterion
$p_{\max} \leq \alpha$.
Under the intersection null~\eqref{eq:H0i} both $p_o$
and $p_r$ are uniformly distributed and so
$p_{\max}$ follows a triangular $\Be(2,1)$ distribution 
\hl{with cumulative distribution function (cdf) 
$F(p)=p^2$,} so {that}
\begin{equation}\label{eq:T1Esquared}
  \Pr(p_{\max} \leq \alpha )
  = \alpha^2 \mbox{ holds for all } \alpha \in (0, 1).
\end{equation}
In the sequel, a $p$-value with this property will be said to have
\emph{exact squared T1E control}. 
The square of a triangular $\Be(2,1)$ distribution follows a 
uniform distribution, so {that} $p = p_{\max}^2=\max\{p_o^2, p_r^2\}$
has cdf
\begin{equation}\label{eq:T1Elinear}
  \Pr(p \leq \alpha 
  ) = \alpha \mbox{ for all } \alpha \in (0, 1).
\end{equation}
In what follows, a $p$-value fulfilling \eqref{eq:T1Elinear}
will be said to have \emph{exact linear T1E
  control}. Note that linear T1E control is the
  traditional requirement for $p$-values \citep[p.~397]{CasellaBerger2002}
  whereas $p$-values with
  squared T1E control (such as $p_{\max}$) are useful if
  the $p$-value is based on two independent studies.


Another way to obtain 
  $p_{\max}$ as the $p$-value from the two-trials rule with squared T1E control is
  by considering~\eqref{eq:equation}, 
  but replacing the 
  variance ratio $c$ by $0$, so $z_S^2 = z_{\min}^2 = \min\{z_o^2, z_r^2\}$.
  Now the distribution 
of  $Y = \min\{z_o^2, z_r^2\}$ under the intersection null 
is of interest. It can be shown that the {random} variable $Y$ has cdf 
\begin{eqnarray}
F_{0}(y) & = & 1 - 4 \left[1-\Phi(\sqrt{y}) \right]^2 \mbox{ for } y \geq 0, 
\label{eq:mincdf} 
\end{eqnarray}
see Supporting Material (SM)~\ref{app:distYmin} for a derivation. 
Now $Y$ doesn't take into account the direction of the effect estimates
and hence the corresponding $p$-value
\begin{eqnarray}\label{eq:4p}
4 \, p = 1- F_0(y=z_{\min}^2) = 4\left[1 - \Phi(\min\{\abs{z_o}, \abs{z_r}\})\right]^2
\end{eqnarray}
is two-sided with exact linear T1E control.  We use the notation $4\, p$, as
there are two studies
(original and replication) with four possible sign combinations of
$\hat \theta_o$ and $\hat \theta_r$. If the
combination~\eqref{eq:direction} is fulfilled,
the one-sided $p$-value is therefore obtained from \eqref{eq:4p} through division 
  by $4$:
\begin{eqnarray*}
  p  =  \left[1-F_0(z_{\min}^2)\right]/4
     =  \left[1 - \Phi(\min\{{z_o}, {z_r}\})\right]^2 
     =  \left(\max\{p_o, p_r\}\right)^2 
     =  \max\{p_o^2, p_r^2\}
\end{eqnarray*}
and so 
$p = p_{\max}^2$ respectively $p_{\max} = \sqrt{p}$. 

\begin{figure}[!h]
\vspace{1.5cm}
  \begin{tikzpicture}[node distance = 4.5cm, auto]
  \node[block2, minimum height = 1.5cm, thick, fill=gray!25] (gen_formula){$p = [1- F_c(z_S^2)]/4$};
      \node[plain2, above = 1.5cm, left of = gen_formula, node distance = 2cm](2tr){$c = 0$ \\ (2TR)}; 
          \node[plain2, below = 1.5cm, left of = gen_formula, node distance = 2cm](pscep) {$c > 0$ \\ (sceptical $p$)}; 
  \node[block, below of = gen_formula](pS){$p_S^*$};
  \node[block2, above of = gen_formula, node distance = 3cm](p2tr){$p_{\max} = \max(p_o, p_r)$};
  \node[block2, right of = gen_formula, above = 1.5cm](2tr_cond){$p \leq \alpha^2 \Leftrightarrow p_{\max} \leq \alpha$};
    \node[block2, right of = gen_formula, below = 1.5cm](ps_cond){$p \leq \alpha^2 \Leftrightarrow p_S^* \leq \alpha$};
    \node[block, right of = ps_cond, above = 1.5cm](rs)(rs){RS};
    
 \path [line] (gen_formula) -- node [midway, right] {$p_{\max} = \sqrt{p}$}(p2tr);
  \path [line] (gen_formula) -- node [midway, right] {$p_S^* = \sqrt{p}$}(pS);
   \path [line] (gen_formula) -- node [midway, right] {}(2tr_cond);
  \path [line] (gen_formula) -- node [midway, right] {}(ps_cond);
     \path [line] (pS) -- node [midway, right] {}(ps_cond);
  \path [line] (p2tr) -- node [midway, right] {}(2tr_cond);
  \path[line] (ps_cond) -- node[midway, above, sloped]{yes}(rs);
  \path[line] (2tr_cond) -- node[midway, above, sloped]{yes}(rs);
  \end{tikzpicture}
  \caption{Replication success (RS) with the two-trials rule (2TR) 
    and the \hlcm{controlled} sceptical $p$-value $p_S^*$. The cdf $F_c(.)$ is based on
    the distribution 
  of $z_S^2$ in \eqref{eq:equation} \hlrev{under the intersection null}
  and the $p$-values are one-sided\hllh{, calculated under the assumption that \eqref{eq:direction} holds}.}
  \label{fig:2tr_rs}
\end{figure}
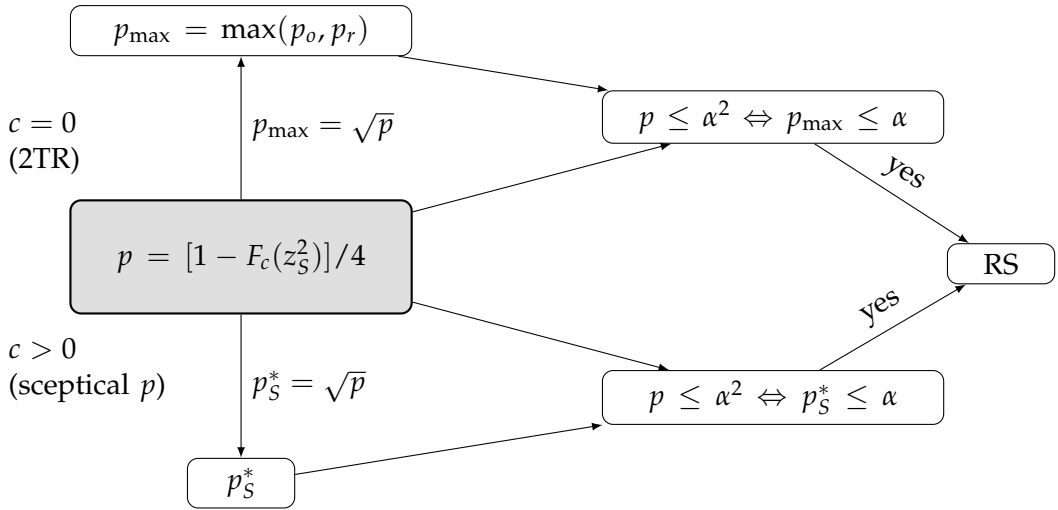

This insight suggests a strategy to obtain a sceptical $p$-value with
exact \hlcm{squared} overall T1E control:
If we can derive the
distribution function $F_c(.)$ of $z_S^2$ in~\eqref{eq:equation} \hlrev{under the intersection null}
for any value of $c>0$,
then we can use the transformation
$p=[1-F_c(z_S^2)]/4$, provided \eqref{eq:direction} holds. 
The 
\emph{controlled} sceptical $p$-value then is
$p_S^* = \sqrt{p}$ and has exact squared \hlcm{overall} T1E control. 
If  \eqref{eq:direction} doesn't hold, we set $p_S^* = 1-\sqrt{p}$.
For simplicity, we call $p_S^*$ the sceptical $p$-value in 
the following, 
if no misunderstandings can arise.
  Figure~\ref{fig:2tr_rs} summarises the different
  steps from the null distribution function $F_c(.)$
   to the assessment of replication success at \hlcm{overall T1E rate $\alpha^2$}
   with the two-trials rule and the sceptical $p$-value. 
  
\subsection{The null distribution for equal variances}\label{sec:harmonic}
We first consider the case $c=1$, where the null distribution of $z_S^2$ is available from the harmonic mean $\chi^2$-test \citep{held2020b}. 
The solution of
\eqref{eq:equation} then is 
\begin{equation}\label{eq:solutionH}
z_S^2 = {z_H^2}/{2}
\end{equation}
where
$z^2_H=2/(1/z_o^2+ 1/z_r^2)$ is the harmonic mean of the squared test
statistics $z_o^2$ and $z_r^2$. 
It \hl{can be shown} that
$z^2_S$
has a gamma $\Ga(1/2, 2)$ distribution under the intersection null \eqref{eq:H0i},
where $z_o^2$ and $z_r^2$ are independent
$\chi^2(1)$-distributed \citep[eq.~(2.3)]{PillaiMeng2016}. 
The cdf $F_{c=1}(y)$ of $Y=z^2_S$ is thus readily
available and a two-sided $p$-value with exact linear T1E control can
be calculated: 
\begin{equation}\label{eq:p1}
  4 \, p = 1-F_{c=1}(y=z_S^2).
\end{equation}
Division by 4 gives the corresponding one-sided $p$-value $p$ 
if~\eqref{eq:direction} is fulfilled
and the square root
$p_S^*=\sqrt{p}$ defines the controlled sceptical
$p$-value with exact squared T1E control for $c=1$.

\subsection{The null distribution for unequal variances}\label{sec:exactT1Econtrol}
For $c > 0$ and $c \neq 1$
there is a unique solution of \eqref{eq:equation} that fulfills the 
requirement $0 \leq z_S^2 \leq \min\{z_o^2, z_r^2\}$:
\begin{eqnarray}
z_S^2 
&=&  
\frac{z_A^2}{c-1} 
  \left\{ \sqrt{1 + (c-1) z^2_H/z_A^2} - 1 \right\},
              \label{eq:solution}
\end{eqnarray}
where 
$z^2_A$ is the arithmetic and $z^2_H$ the harmonic
mean of $z_o^2$ and $z_r^2$.  
To obtain $F_c(.)$,
consider the probabilistic version of equation \eqref{eq:solution}, where
the random variable $Y=z_S^2$ depends on the two
random variables $z_o^2$ and $z_r^2$ through $z_A^2$ and $z_H^2$.
Under the intersection null hypothesis~\eqref{eq:H0i},
$z_o^2$ and $z_r^2$ are independent
$\chi^2(1)$-distributed. Then $z_A^2$ and $z_H^2/z_A^2$ in \eqref{eq:solution} 
are also independent \citep[Section 4.7]{GrimmettStirzaker2001}, 
which facilitates the computation of the  
cdf $F_c(y) = \Pr(Y \leq y \given c)$ of $Y$. 
In SM~\ref{app:distY} we show that
\begin{equation}\label{eq:Fc}
  F_c(y) = 1 - \frac{1}{\pi} \int_0^{1} 
    \frac{\exp\{-g(y, t, c)\}}{\sqrt{t(1-t)}} \, dt 
\end{equation}  
where
\[
  g(y, t, c) = 
  \frac{{(c-1)y}}{{\sqrt{1 + (c-1) t}-1}}.
\]
\begin{figure}[ht!]
\begin{knitrout}
\definecolor{shadecolor}{rgb}{0.969, 0.969, 0.969}\color{fgcolor}
\includegraphics[width=\maxwidth]{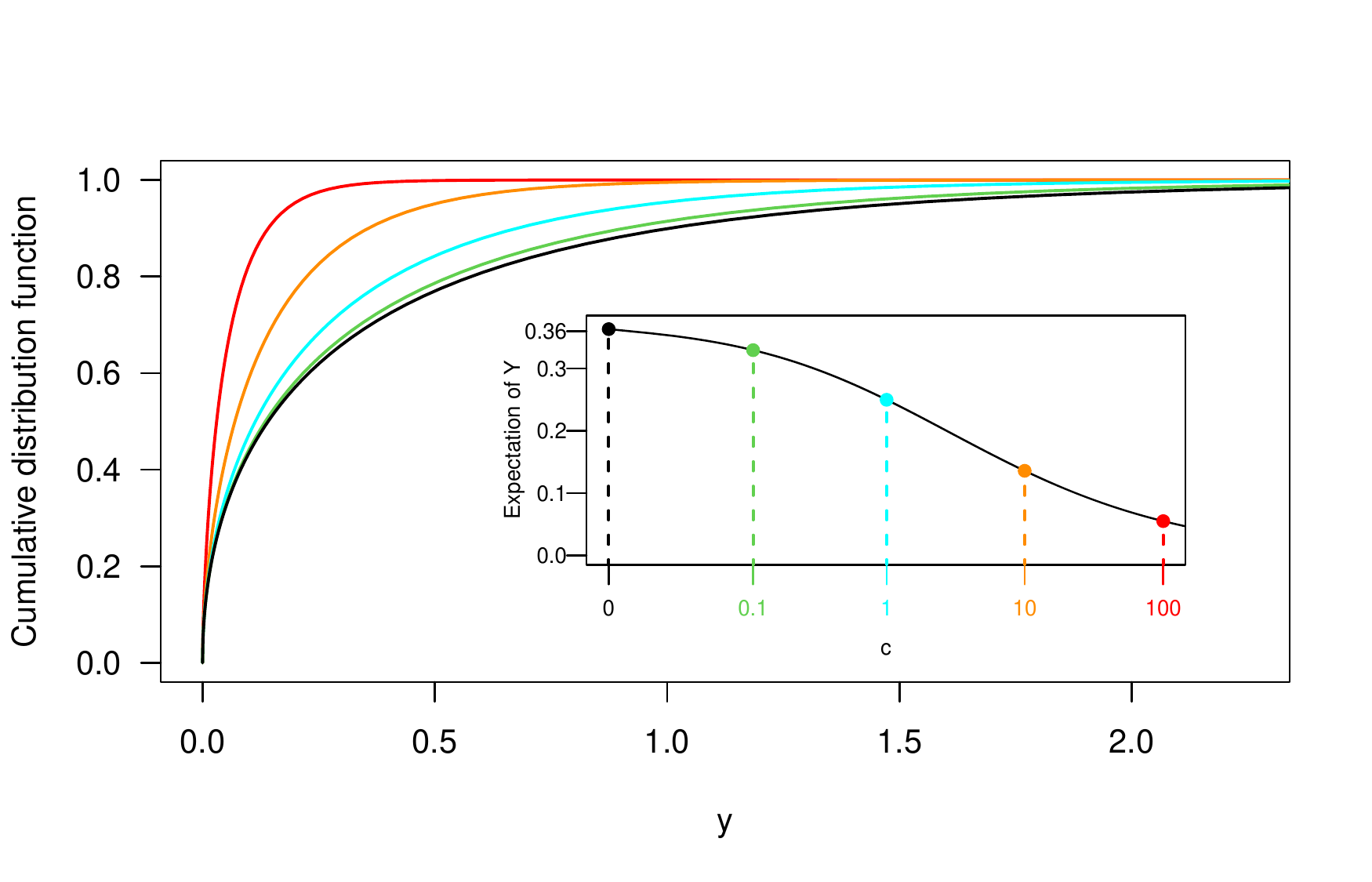} 
\end{knitrout}
\vspace{-0.6cm}
\caption{\label{fig:cdf} Cumulative distribution function 
$F_c(y)$ of $Y=z_S^2$  (main plot) and expectation of $Y$ (inset plot) under 
the intersection null hypothesis~\eqref{eq:H0i} for
  different values of $c$.
  }
\end{figure}

Figure \ref{fig:cdf} compares $F_c(y)$ and the expectation of $Y$
for different values of $c$, including the special cases $c=0$ and $c=1$.
Evaluation of \eqref{eq:Fc} is possible with numerical
integration techniques and so
a two-sided  $p$-value with exact linear T1E control can be calculated:
\begin{equation}\label{eq:p2}
  4 \, p = 1-F_c(z_S^2)
\end{equation}
with $z_S^2$ as defined in \eqref{eq:solution}. If \eqref{eq:direction} is fulfilled, 
the corresponding one-sided $p$-value
has exact linear T1E control 
and $p_S^*=\sqrt{p}$ defines the one-sided controlled sceptical
$p$-value with exact squared T1E
control.
\subsection{Limiting cases}\label{sec:limitingcases}
For $c \downarrow 0$ the \hl{two-sided} $p$-value $4 \, p$ in
\eqref{eq:p2} 
converges to $4 \, p^2_{\max}$. This follows from the fact that
$z_S^2 \uparrow z^2_{\min}$ for $c \downarrow 0$
\cite[eq.~(11)]{held2020} with cdf \eqref{eq:mincdf} under the intersection null.
The two-trials rule described in Section \ref{sec:2TR} is therefore a limiting case of our framework if we are 
willing to ignore the interpretation of $c$ as the variance ratio.

For $c \rightarrow \infty$ the two-sided 
$p$-value $4 \, p$ in \eqref{eq:p2} converges to 
\begin{equation}\label{eq:limp}
4 \, p_\infty = \lim_{c \rightarrow \infty} 4 \, p = \frac{1}{\pi} \int_0^{1} 
  \frac{\exp\left(- z_G^2 / \sqrt{t}\right)}{\sqrt{t(1-t)}} \, dt ,
\end{equation}
where $z_G^2 = \sqrt{z_o^2 z_r^2} = \abs{z_o z_r}$ is the geometric mean of the squared test
statistics $z_o^2$ and $z_r^2$ (proof to be found in SM~\ref{app:plimit}).
Note that $4 \, p_\infty = 1$ if either $z_o=0$ or $z_r=0$, as
$f(x)=1/\{\pi \sqrt{x(1-x)}\}$ is the density of a $X \sim \Be(1/2,1/2)$ random variable and integrates to $1$.
Furthermore, \eqref{eq:limp} is a two-sided $p$-value
with exact linear T1E control, \ie $4 \, p_\infty$
is uniformly distributed on the unit interval
if $z_o^2$ and $z_r^2$ are i.i.d.~$\chi^2(1)$, see SM~\ref{app:uniform} for a proof. 

Other well-known methods that combine two (or more) $p$-values
 are Fisher's \hlrev{\citep{Fisher1958}}, Stouffer's \hlrev{\citep{Stouffer1949}}
 and Pearson's \hlrev{\citep{Pearson1933}} methods.
Fisher's method is based on the product of 
the $p$-values, Pearson's method on the product of one minus the $p$-values,
Stouffer's method on the sum of the $z$-values, whereas \eqref{eq:limp} is based
on the product of the $z$-values. We will compare the different methods to
combine $p$-values in Sections~\ref{sec:adaptiveLevel} 
and~\ref{sec:combTest} in more detail.

\subsection{Properties of the sceptical $p$-value}\label{sec:comparison}



\begin{figure}[!h]
\centering
\begin{knitrout}
\definecolor{shadecolor}{rgb}{0.969, 0.969, 0.969}\color{fgcolor}
\includegraphics[width=\maxwidth]{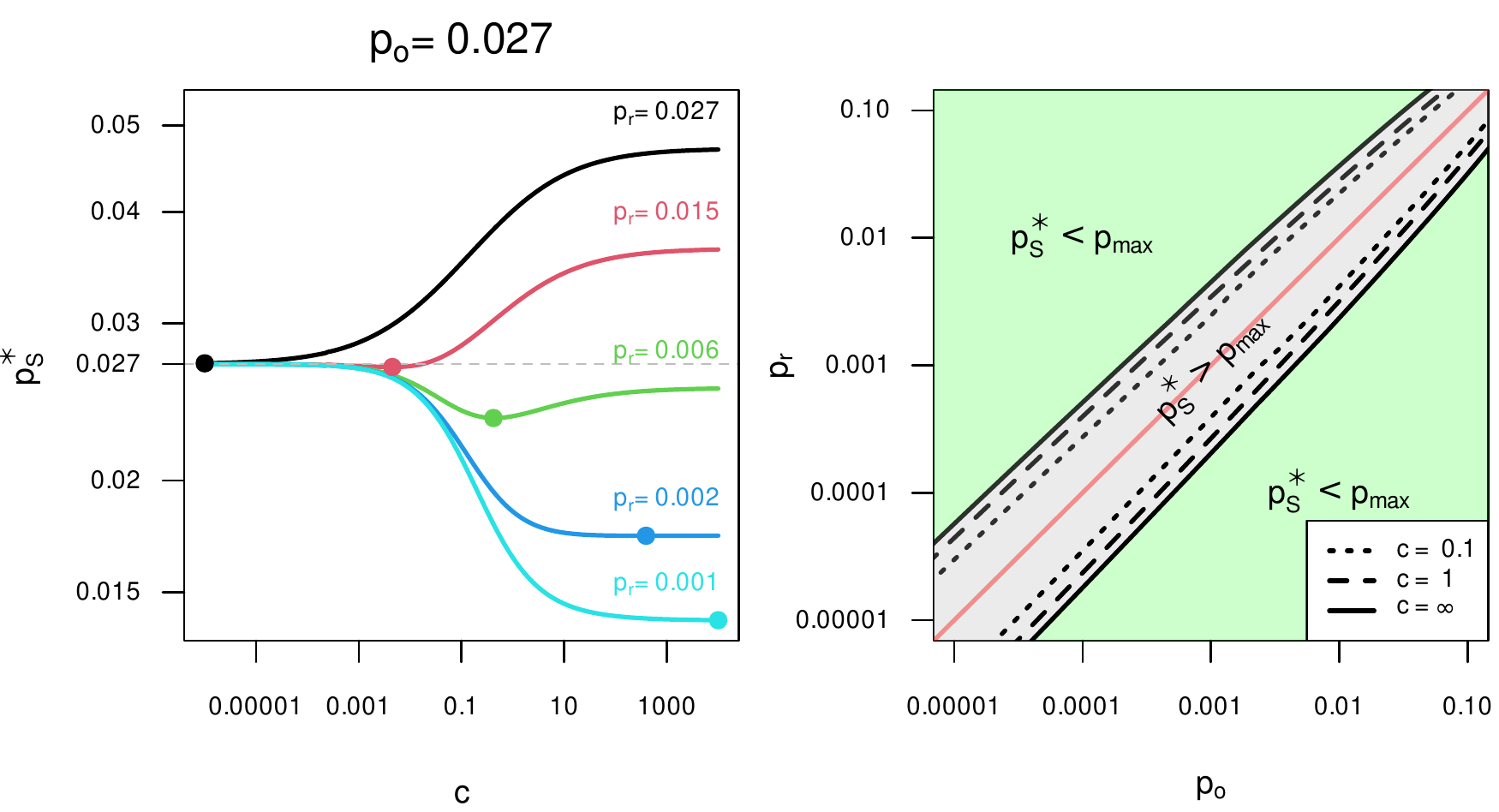} 
\end{knitrout}
\caption{Left panel: Sceptical $p$-value $p_S^*$ as a function of the variance 
  ratio $c$ for $p_o = 0.027$ and selected values of $p_r$. 
The dots represent 
the infimum of $p_S^*$ for selected values of $p_r$ and the dashed line 
indicates $p_{\max}$.
Right panel: Combinations of $p_o$ and $p_r$ for which $p_S^*$ is always smaller 
than $p_{\max}$ (green area), for which $p_S^*$ is always larger than 
$p_{\max}$ (red line) and for which it depends on the variance ratio $c$ (gray region). 
In the gray region above (respectively below) the red line, $p_S^*$ is smaller than 
$p_{\max}$ for combinations of $p_o$ and $p_r$ laying above (respectively below)
the black line for the corresponding $c$, 
and vice-versa. 
}
\label{fig:versus}
\end{figure}

For fixed $p_o$ and $p_r$, the nominal and golden versions of the sceptical 
$p$-value monotonically increase with increasing 
variance ratio $c$ \citep[Section 3.1]{held2020} 
\hlrev{and eventually reach $0.5$ for $c \rightarrow \infty$}. 
The controlled sceptical $p$-value $p_S^*$ behaves differently, as illustrated
in Figure~\ref{fig:versus} (left), which shows $p_S^*$ as a function of $c$ for
selected values of $p_o$ and $p_r$.

As described in Section~\ref{sec:limitingcases}, 
  $p_S^*$ converges to $p_{\max}$ for $c \downarrow 0$. The functional behaviour of
    $p_S^*$ as a function of $c$ can be studied through inspection of the 
derivative of $p_S^*$ with respect to $c$, see
SM~\ref{sec:deriv} and~\ref{sec:monotone}.
If $p_o = p_r$ then $p_S^*$ increases monotonically 
with increasing $c$. If the difference between $p_o$ 
and $p_r$ is relatively large, $p_S^*$ decreases monotonically.
If $\abs{p_o - p_r}$ is relatively small but not zero, 
the sceptical $p$-value first 
decreases and then increases with increasing $c$. 
The infimum of $p_S^*$ is hence
$p_{\max}$ in the first case, $p_{\infty}$ from~\eqref{eq:limp}
in the second, and in between those two
  values in the third case.

These properties allow us to compare $p_S^*$ and $p_{\max}$ for
  the same values of $p_o$ and $p_r$, see Figure~\ref{fig:versus} (right).  In the
  first case ($p_o = p_r$), $p_S^*$ is always larger than $p_{\max}$,
  for any value of the variance ratio $c$. If $p_o$ and $p_r$ differ
  considerably, $p_S^*$ is always smaller than $p_{\max}$ (the green
  region). The gray area 
  depicts combinations of $p_o$ and $p_r$ for
  which the sceptical $p$-value $p_S^*$ is not monotone as a function of $c$,
  but first decreases, then increases and eventually gets larger than
  $p_{\max}$ for large $c$.  Whether $p_S^*$ is smaller or larger than
  $p_{\max}$ now depends on the value of $c$, as indicated with
  dotted and dashed lines in Figure~\ref{fig:versus} (right) for $c=0.1$ and
  $c=1$, respectively.

\section{\hl{Replication success rates and regions}}\label{sec:rs}

In this section, we consider properties of the sceptical $p$-value
based on the dichotomous criterion $p_S^* \leq \alpha$ for replication success. 
  We start in Section \ref{sec:adaptiveLevel} with
deriving the corresponding value of \hlcm{the replication success level} $\gamma$
in \eqref{eq:general}  
  so that the overall T1E rate is exactly $\alpha^2$ for a particular value of $c > 0$.
  As discussed in Section \ref{sec:null_hyp},  the replication success level 
  $\gamma$ is \hlcm{also} a bound on the partial T1E rate of the sceptical $p$-value.
We will then compare success regions for different values of $c$ in Section \ref{sec:combTest} and investigate project power in Section \ref{sec:pp}. Finally, Section \ref{sec:design} outlines how the sceptical $p$-value can be used for sample size calculations.
This is facilitated through the interpretation of the variance ratio $c$ as relative sample size, $c=n_r/n_o$,
  since the variances of the effect estimates 
are usually inversely proportional to the corresponding sample sizes $n_o$
and $n_r$, \ie $\sigma_o^2 = \kappa^2/n_o$ and
$\sigma_r^2 = \kappa^2/n_r$ for some unit variance $\kappa^2$ \citep{held2020}.

\subsection{Partial Type-I error control}\label{sec:adaptiveLevel}

\begin{table}
\begin{center}
  \begin{tabular}{llrr}
  Method & Parameters & \multicolumn{2}{c}{T1E control} \\ \hline
         && overall & partial \\
  \hline 
  Two-trials rule & $p_o$, $p_r$ & $ = \alpha^2$ & $< \alpha$ \\
  Fisher & $p_o$, $p_r$ & $ = \alpha^2$ & \hlrev{$< 1$} \\
  Stouffer & $p_o$, $p_r$ & $ = \alpha^2$ & \hlrev{$< 1$} \\
  \hlrev{Pearson} & \hlrev{$p_o$, $p_r$} & \hlrev{$ = \alpha ^2$} &  \hlrev{$< c_P(\alpha)$} \\
  nominal $p_S$ &  $p_o$, $p_r$, $c$ & $< \alpha^2$ & $< \alpha$ \\
    golden $\tilde p_S$ &  $p_o$, $p_r$, $c$ & $< \gamma(\alpha)^2$ 
         & $< \gamma(\alpha)$ \\
  controlled $p_S^*$ &  $p_o$, $p_r$, $c$ &  $= \alpha^2$ & $<\gamma_c(\alpha)$ \\
\end{tabular}
\caption{T1E rate of different methods to assess replication success. The overall T1E rate is calculated
  under the intersection null \eqref{eq:H0i}. The partial T1E rate
  is calculated under the union null 
  \eqref{eq:H0c}.   
  \label{tab:T1E}
  }
\end{center}
\end{table}

If we want to know the bound on the partial T1E rate of the controlled sceptical $p$-value, 
we need to derive the corresponding value of
$\gamma$ in \eqref{eq:general}, now denoted as $\gamma_c = \gamma_c(\alpha)$ 
as it depends not only on the target overall T1E rate
$\alpha^2$, but also the relative sample size
$c$. \hlcm{Comparing  $p_S$ to $\gamma_c(\alpha)$ is then equivalent to comparing $p_S^*$ to $\alpha$.}

For $c=1$, we have
 ${\gamma_1} = 1-\Phi(\Phi^{-1}(1-2 \alpha^2)/2)$, 
see \citet[Section 2.1]{held2020b}.
For example, 
$\gamma_1(\alpha = 0.025) = 0.065$, so the partial T1E rate
is bounded by 0.065 for $c=1$.
The null distribution function~\eqref{eq:Fc} of $z_S^2$
can be used to compute the bound $\gamma_c$ for $c \neq 1$,
but now numerical methods are needed.
Briefly, the overall T1E rate for any two values of
$c$ and $\gamma_c$ can be
computed with numerical integration \citep[Section 3.2]{held_etal2020}.  Root-finding methods are then used to find the
value of $\gamma_c$ which gives the target overall T1E rate of $\alpha^2$.
The inset plot in Figure \ref{fig:fig3b} shows the bound $\gamma_c$
as a function of $c$ for exact overall T1E control at 
$\alpha^2=0.000625$.
\hlrev{The bound on the partial T1E rate can get quite large for large relative 
sample sizes $c$. For 
example, it is $\gamma_{10}(\alpha = 0.025) = 0.14$ for $c = 10$. 
However, this large partial T1E rate is balanced by the fact that such large 
relative sample sizes 
are only needed for unconvincing original studies, where success is unlikely. 
The conditional T1E rate for such unconvincing original studies 
combined with large relative sample sizes is in fact very small, 
as shown in Section~\ref{sec:design}.}

A summary of the overall and partial 
T1E rates of different methods is given in Table \ref{tab:T1E}. 
Fisher's, Stouffer's \hlrev{and Pearson's} methods control 
the overall T1E rate exactly
at \hlrev{significance} level $\alpha^2$. However, 
\hlrev{as the first two methods} do not impose 
a threshold on the individual $p$-values $p_o$ and $p_r$ \citep{Rosenkrantz2002}, 
the partial T1E rate is bounded \hlrev{by one}
and replication success can occur if one of 
the two $p$-values is very large.
\hlrev{The partial T1E rate of Pearson's method is bounded by 
 $c_P(\alpha) = 1 - \exp(-0.5 \chi^2_4(\alpha^2))$, for example 
 $c_P(\alpha = 0.025) = 0.035$.}

\subsection{Success regions}\label{sec:combTest}

\begin{figure}

\begin{knitrout}
\definecolor{shadecolor}{rgb}{0.969, 0.969, 0.969}\color{fgcolor}
\includegraphics[width=\maxwidth]{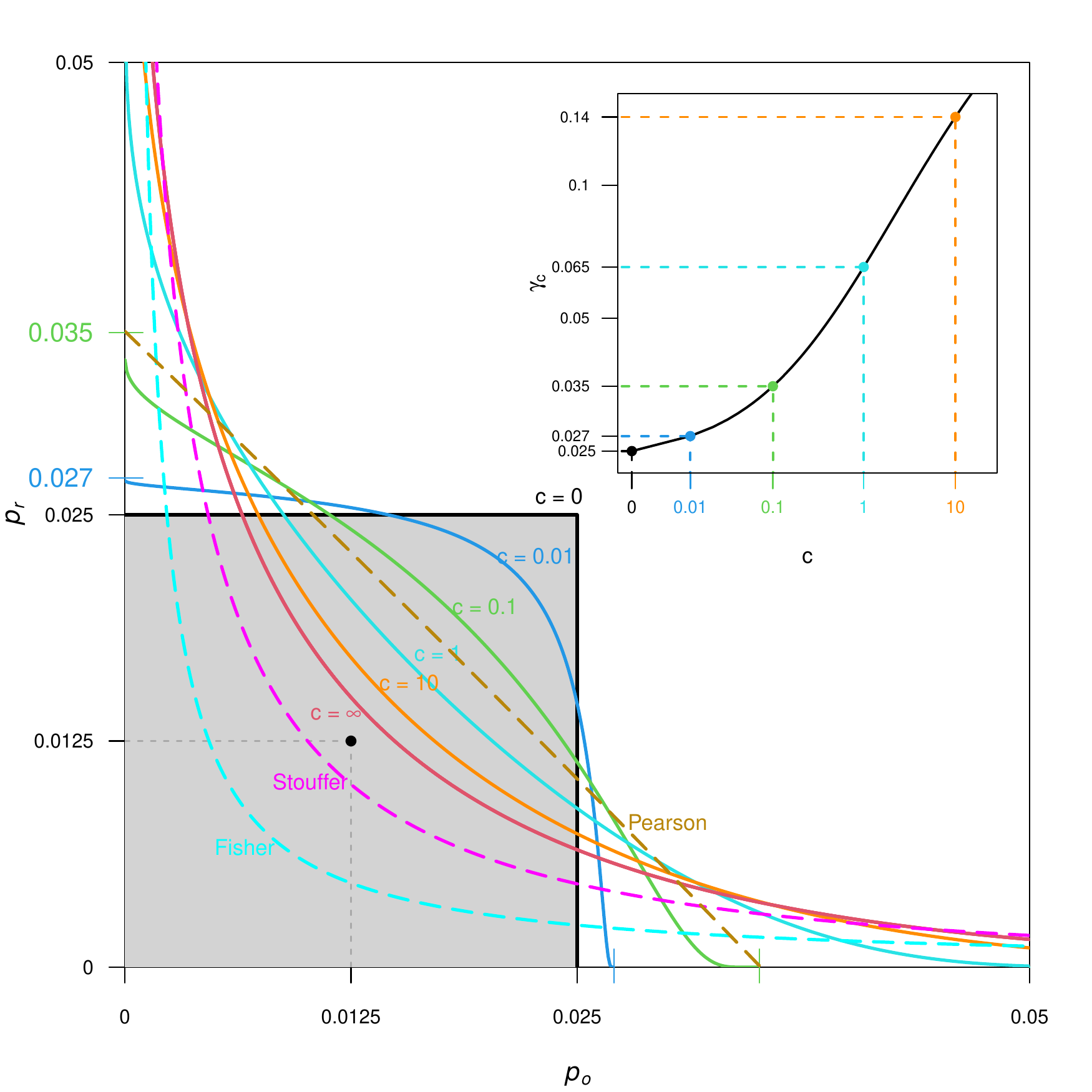} 
\end{knitrout}
\vspace{-0.2cm}
\caption{Success region of the sceptical $p$-value as a function of $p_o$ and $p_r$
  for different 
values of the relative sample size $c$. 
The colored labels on the $y$-axis are the different values of \hllh{the bound} $\gamma_c$
\hlrev{on the partial T1E rate}
($\gamma_1=0.065$ and
  $\gamma_{10}=0.14$ are outside the axis range, but can be read off the inset plot).
The two-trials rule success region is the squared gray area below the black 
  line where $c=0$ and $\gamma_0 = \alpha$. 
Fisher's, 
Stouffer's \hlrev{and Pearson's} methods have been added for comparison purposes.
All methods control the overall T1E 
  rate at $\alpha^2=0.025^2 = 0.000625$, 
  and so the area under each curve is equal to this value. }
  \label{fig:fig3b}
\end{figure}

The (one-sided) framework \eqref{eq:general} can
now be used to determine the region where two
$p$-values $p_o$ and $p_r$ lead to replication success.
The main plot in Figure \ref{fig:fig3b} compares the success regions 
for $\alpha = 0.025$ and
selected values of the variance ratio (respectively relative sample size) $c$, so the
area under each curve is equal to $\alpha^2 = 0.000625$.
Each success region is bounded on both axes by the corresponding
success level $\gamma_c$.
The two-trials rule success region corresponds to
$\gamma_0=0.025$ and is the squared
gray area below the black line.
The success region of the sceptical $p$-value
is close to the two-trials rule's success region for small $c$,
but becomes more and more in favour of $p$-values of different size
as $c$ increases. The case $c = \infty$ is
based on the one-sided $p$-value $p_\infty$ available from
\eqref{eq:limp}.
Also shown are the success regions based on
Fisher's, Stouffer's \hlrev{and Pearson's} method. \hlrev{The first two} are even less in favour of 
$p$-values of the same magnitude.
For example, if both
$p$-values are equal to $\alpha/2=0.0125$ (the solid black point in Figure \ref{fig:fig3b}), replication
success will be flagged with the sceptical $p$-value for any value of $c$, but not with
Fisher's nor Stouffer's method.
\hlrev{Pearson's method has a success region similar to that of
the sceptical $p$-value with $c = 0.1$, and the same bound on the partial 
T1E rate.}

\subsection{Project power}\label{sec:pp}

Suppose none of the two studies has been conducted yet and
so the probability to declare replication success is calculated
over both studies in combination for a fixed relative sample size $c$.
Using numerical integrations adapted from \citet[Section 3.3]{held_etal2020}, 
the project power of the sceptical $p$-value 
is considered in this section.

\begin{figure}[!ht]

\begin{knitrout}
\definecolor{shadecolor}{rgb}{0.969, 0.969, 0.969}\color{fgcolor}
\includegraphics[width=\maxwidth]{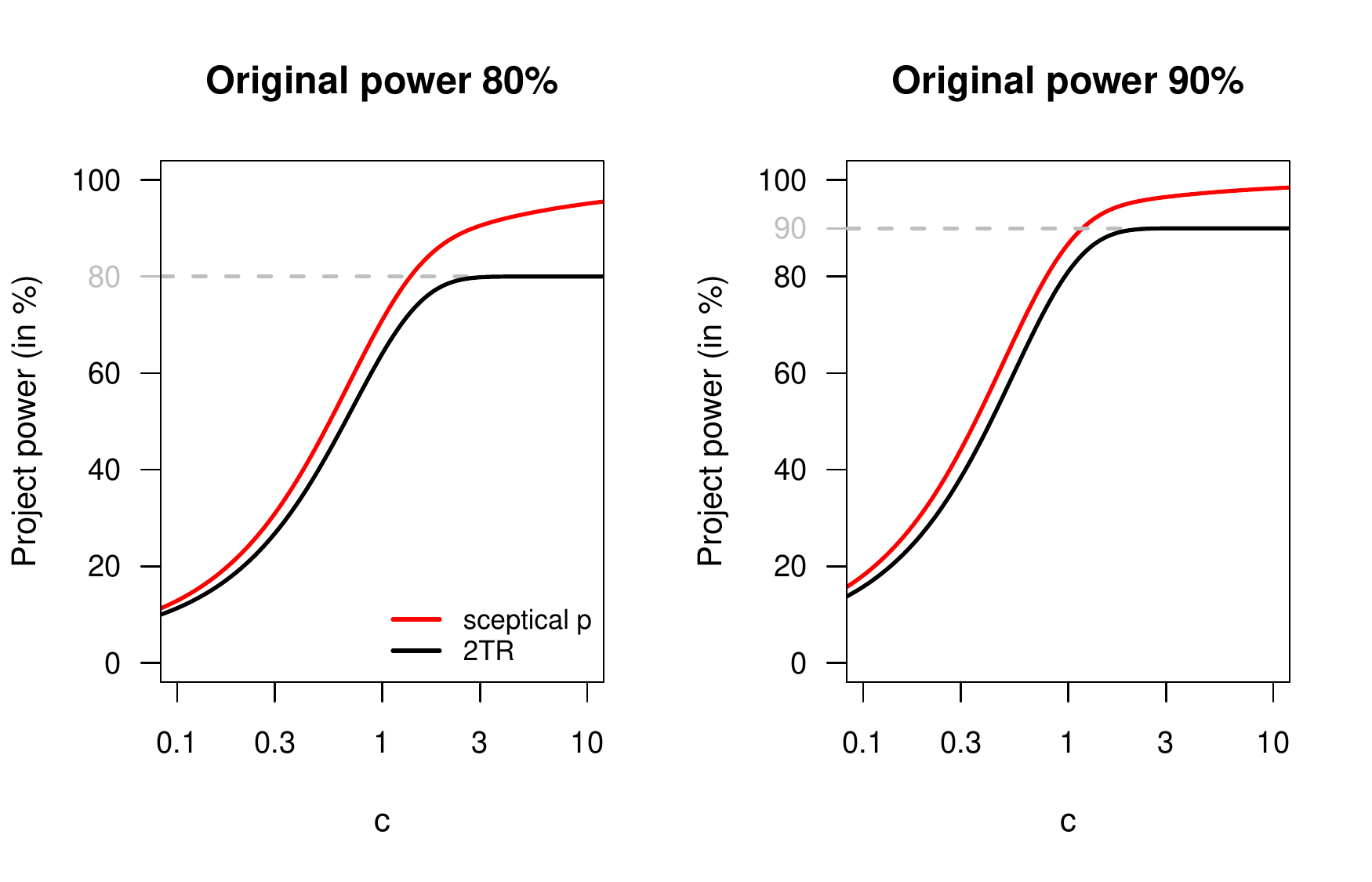} 
\end{knitrout}
\vspace{-0.2cm}
\caption{\label{fig:fig5}  Project power
as a function of the relative sample size $c$ for $\alpha = 0.025$.
  The original power is 80\% in the left plot, and 90\%
  in the right one. 
Results are given for the sceptical $p$-value and
  compared with the two-trials rule (2TR).}
\end{figure}

The project power is the probability to declare replication success 
when both effects are equal and non-null. 
The distribution of $z_o$ is then 
$\Nor(\mu = z_\alpha + z_\beta, 1)$, where $1- \beta$ is the power
to detect the true original effect $\theta_o = \mu\, \sigma_o$
\citep[Section 3.3]{mat2006}, 
and the distribution of $z_r$ is $z_r\sim \Nor(\sqrt{c} \mu, 1)$.
Figure~\ref{fig:fig5} shows the project power 
of the sceptical $p$-value and the two-trials rule with $\alpha = 
0.025$ and original power $1 -  \beta = 80$\% (left), respectively $90$\% (right).
The project power based on the two-trials rule
converges to 80\%, respectively 90\%,
for large relative sample size $c$.  
The project power based on the sceptical $p$-value  is 
always larger than with the two-trials rule, 
and increases to values close to 100\% for large $c$. 
For example, for 80\% original power and $c=2$ the project power of the
two trials rule is 78\%, while the
project power of the
sceptical $p$-value is already 87\%.

\subsection{\hl{Design of replication studies}}\label{sec:design}

In this section we assume that the original study has already been 
conducted and a replication study is planned. 
It has been recently argued that the method used for sample size planning should 
always match the one used for the analysis \citep{Anderson2022}.
Hence, we develop methods to
design the replication study based on the sceptical $p$-value 
and compare it to the design based on the two-trials rule, 
\ie significance of the replication study.

The probability to declare replication success with
a particular sample size $n_r$ is known as
the power of the replication study and is often calculated conditional on 
the effect estimate from the original study.
Predictive power \citep{SpiegelhalterFreedman1986} can also be used and takes
the uncertainty of the original effect estimate into account.
Formulas for the power of the two-trials rule and the sceptical 
$p$-value at fixed success level can be found 
in \citet[Section 2.1]{MicheloudHeld2022} and \citet[Section 3.1]{held_etal2020}, 
respectively.

\begin{figure}[!ht]
\begin{knitrout}
\definecolor{shadecolor}{rgb}{0.969, 0.969, 0.969}\color{fgcolor}
\includegraphics[width=\maxwidth]{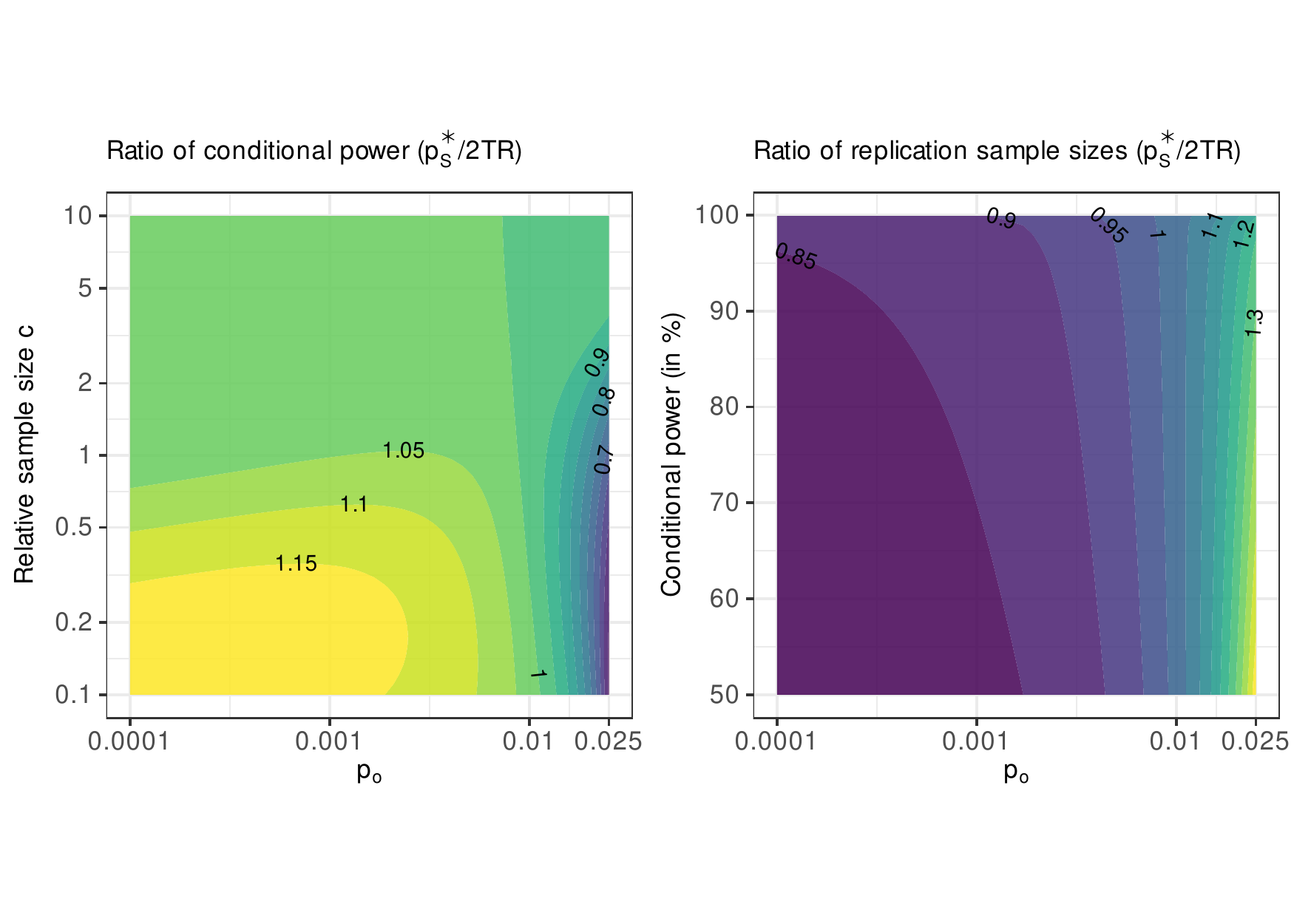} 
\end{knitrout}
\vspace{-1cm}
\caption{
\hl{Ratio of conditional power (left) and replication 
sample sizes (right) calculated with  the
   sceptical $p$-value ($p_S^*$) versus the two-trials rule (2TR) as a function 
  of the original $p$-value $p_o$ and the relative sample size 
  $c$ (left) or the conditional power (right) for $\alpha = 0.025$.}
  }
\label{fig:powRep}
\end{figure}

Figure~\ref{fig:powRep} (left) shows the ratio of 
conditional power calculated with the
sceptical $p$-value versus the two-trials rule with $\alpha = 0.025$ 
as a function of the relative sample size $c$ and the original 
$p$-value $p_o$. 
The sceptical $p$-value has larger power (ratio > $1$)
if the original study is already convincing.
For example, for $c = 1$, this is the case if
$p_o < 0.01$,
otherwise the two-trials rule has larger power.
However, if $p_o > \alpha$, the power of the two-trials rule is $0$, 
but not the power of the sceptical $p$-value as long as $p_o < \gamma_c$.



\begin{figure}[!h]
  \centering
\begin{knitrout}
\definecolor{shadecolor}{rgb}{0.969, 0.969, 0.969}\color{fgcolor}
\includegraphics[width=\maxwidth]{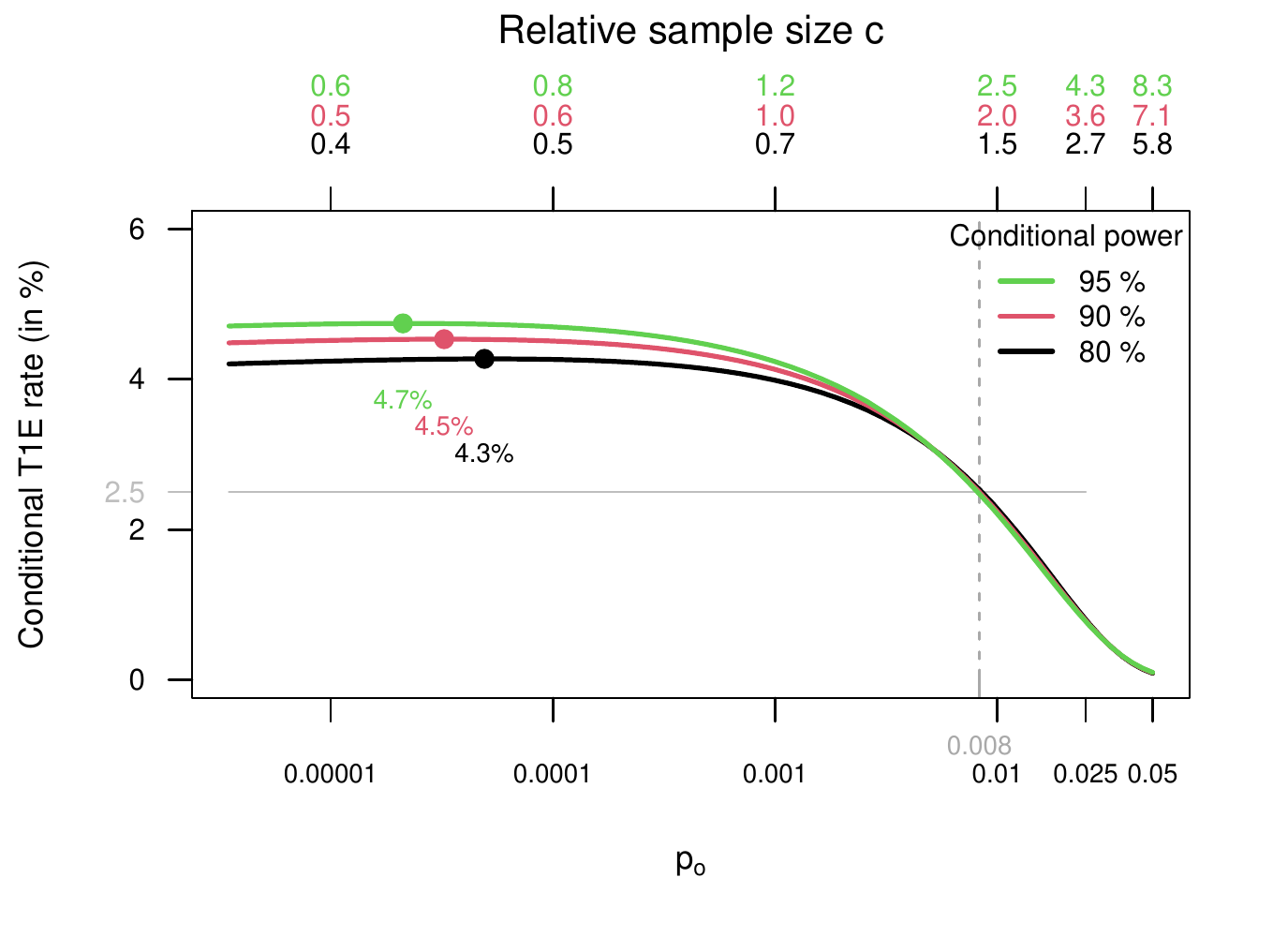} 
\end{knitrout}
\caption{Conditional T1E rate of the sceptical $p$-value
as a function of the original $p$-value
$p_o$. 
The relative sample size (top axis) is calculated with the sceptical $p$-value method 
to reach a conditional power of 80\%, 90\%, and 95\% 
\hlcm{at $\alpha = 0.025$}. 
Each dot represents the upper bound for the conditional T1E rate with the 
respective power.
The gray horizontal line indicates the T1E rate of the two-trials rule.}
\label{fig:rep_t1e}
\end{figure}

Instead of calculating the power for a fixed replication sample size 
$n_r$, we can also fix the power to a desired value and 
calculate the required sample size $n_r$.
Sample size calculation with the controlled
sceptical $p$-value does not have a closed-form expression, because the 
success level
$\gamma_c$ in \eqref{eq:general} depends on the relative sample size $c$.
Root-finding algorithms are therefore required to find the value of $c$
which leads to the desired power. Importantly, sample size calculation is
  now possible even for non-significant original studies, as shown in the top 
  axis of Figure \ref{fig:rep_t1e} for conditional power values of 80, 90, and 
  95\%. Note that the required relative sample size can become quite large
  if $p_o > \alpha$.

Figure~\ref{fig:powRep} (right panel) shows the ratio of the replication sample sizes 
  calculated with the sceptical $p$-value versus the two-trials rule.
This is done only for significant original studies, as this is required for success with the two-trials rule.
The sceptical $p$-value requires less samples
than the two-trials rule 
for already convincing original studies ($p_o < 0.007$).
Similar results are obtained when predictive power is used instead 
of conditional, see SM~\ref{sec:predPower}.

The T1E rate of the sceptical $p$-value can also be considered 
under $H_0^{\, r}$, conditional on the original study result. First, the
relative sample size $c$ to reach a certain power for a 
fixed original $z$-value $z_o$ is calculated. These values of 
$z_o$ and $c$ are then used in~\eqref{eq:general} to derive a lower 
bound for the replication $z$-value $z_r$
to achieve replication success:
\begin{equation}\label{eq:zLower}
z_r \geq z_{\gamma_c} \sqrt{1+c/(z_o^2/z_{\gamma_c}^2-1)}.
\end{equation}
Subsequent transformation 
of the right hand side of \eqref{eq:zLower} to the corresponding upper bound 
for $p_r$ gives the conditional T1E rate.
Note that the conditional T1E rate of the two-trials rule is 
constant at $\alpha$ as long as $p_o \leq \alpha$.
Figure~\ref{fig:rep_t1e} shows the conditional T1E
  rate of the sceptical $p$-value as a function of the $p$-value $p_o$ from the original study.
The relative sample size $c$ in \eqref{eq:zLower} is calculated with the sceptical $p$-value method 
to reach a conditional power of 80\%, 90\%, and 95\% respectively
\hlcm{with $\gamma_c(\alpha = 0.025)$}.
The conditional T1E rate is 
  larger than $\alpha = 2.5$\%, 
 the conditional T1E rate of the two-trials rule, for $p_o < 0.008$ but
  bounded by 
  $4.3$\%, $4.5$\% and 
  $4.7$\% for a power of 
   80\%, 90\% and 95\%, respectively. 
   \hlrev{The conditional T1E rate is hence never larger than 
   2$\alpha = 5\%$, 
   and this also holds for other values of $\alpha$ provided that 
   they are not very small, see SM~\ref{sec:predPower}.
   }
If $p_o > 0.008$, the conditional T1E rate of the sceptical $p$-value is smaller than  
 $2.5$\% in all three cases.
This illustrates that the
conditional T1E rate is sufficiently bounded if the replication sample size is computed 
based on standard power values to detect the observed effect  from the original study.

A similar pattern can be seen if the conditional T1E rate is based on
predictive rather than conditional power, see
SM~\ref{sec:predPower}, where we also show the conditional T1E rate of the 
nominal and golden sceptical $p$-values. 

\section{Application}\label{sec:appEE}


We now illustrate the proposed methodology using all 18
pairs of studies from the Experimental Economics Replication Project
\citep[EERP]{Camerer2016}.  The different effect estimates were all transformed to correlation
coefficients, where Fisher's $z$-transformation achieves
asym\-ptotically normal effect estimates $\hat \theta_i$ with known standard
errors. 
Table~\ref{tbl:EE_studies} summarizes the results for 
each of the 18 studies.
\subsection{Comparison of two-trials rule versus sceptical $p$-value}

The last two columns in Table~\ref{tbl:EE_studies} show the $p$-value $p_{\max}$
from the two-trials rule and the sceptical $p$-value $p_S^*$, 
respectively. 
There is generally a good agreement, with
$p_S^* < p_{\max}$ for 12 out of the 18 studies.

\begin{table}[!h]
\resizebox{1.1\textwidth}{!}{
\begingroup\small
\begin{tabular}{cccccccccc}
  \toprule
Study & $\hat\theta_o$ & $\hat\theta_r$ & $p_o$ & $p_r$ & Power ($\%$) & $c$ & $c^\star$ & $p_{\max}$ & $p_S^*$ \\ 
  \rowcolor[gray]{0.9} \midrule
de Clippel et al. (2014) & 0.12 &  0.27 & 0.0005 & < 0.0001 & 91.0 & 1.0 & 0.8 & 0.0005 & < 0.0001 \\ 
  Kogan et al. (2011) & 0.34 &  0.31 & < 0.0001 & 0.0005 & 93.9 & 0.7 & 0.6 & 0.0005 & 0.0002 \\ 
   \rowcolor[gray]{0.9}Fudenberg et al. (2012) & 0.31 &  0.34 & 0.0003 & < 0.0001 & 93.9 & 1.0 & 0.9 & 0.0003 & 0.0003 \\ 
  Dulleck et al. (2011) & 0.91 &  0.93 & < 0.0001 & 0.0004 & 90.8 & 0.7 & 0.6 & 0.0004 & 0.0003 \\ 
   \rowcolor[gray]{0.9}Friedman and Oprea (2012) & 0.76 &  0.47 & < 0.0001 & 0.002 & 99.6 & 0.5 & 0.4 & 0.002 & 0.0003 \\ 
  Bartling et al. (2012) & 0.91 &  0.79 & 0.003 & 0.0006 & 96.3 & 1.9 & 1.7 & 0.003 & 0.003 \\ 
   \rowcolor[gray]{0.9}Kessler and Roth (2012) & 0.53 &  0.36 & < 0.0001 & 0.008 & 94.6 & 0.2 & 0.1 & 0.008 & 0.003 \\ 
  Charness and Dufwenberg (2011) & 0.40 &  0.38 & 0.005 & 0.001 & 89.0 & 1.6 & 1.5 & 0.005 & 0.005 \\ 
   \rowcolor[gray]{0.9}Kirchler et al (2012) & 0.80 &  0.60 & 0.008 & 0.005 & 93.7 & 2.1 & 2.1 & 0.008 & 0.011 \\ 
  Fehr et al. (2013) & 0.49 &  0.32 & 0.006 & 0.013 & 92.3 & 1.8 & 1.7 & 0.013 & 0.015 \\ 
   \rowcolor[gray]{0.9}Ambrus and Greiner (2012) & 0.32 &  0.23 & 0.027 & 0.006 & 93.3 & 3.2 & 4.2 & 0.027 & 0.024 \\ 
  Ericson and Fuster (2011) & 0.22 &  0.12 & 0.015 & 0.027 & 92.3 & 2.4 & 2.7 & 0.027 & 0.032 \\ 
   \rowcolor[gray]{0.9}Huck et al. (2011) & 1.19 &  0.39 & 0.0002 & 0.082 & 99.1 & 1.4 & 1.2 & 0.082 & 0.045 \\ 
  Abeler et al. (2011) & 0.18 &  0.08 & 0.023 & 0.08 & 90.7 & 2.7 & 3.4 & 0.08 & 0.074 \\ 
   \rowcolor[gray]{0.9}Chen and Chen (2011) & 1.23 &  0.17 & 0.017 & 0.28 & 98.3 & 3.7 & 4.1 & 0.28 & 0.24 \\ 
  Ifcher and Zarghamee (2011) & 0.29 & -0.01 & 0.016 & 0.53 & 90.7 & 2.3 & 2.6 & 0.53 & 0.53 \\ 
   \rowcolor[gray]{0.9}Duffy and Puzzello (2014) & 1.00 & -0.12 & 0.007 & 0.66 & 95.0 & 2.2 & 2.1 & 0.66 & 0.69 \\ 
  Kuziemko et al. (2014) & 0.29 & -0.12 & 0.035 & 0.92 & 93.1 & 3.6 & 5.3 & 0.92 & 0.92 \\ 
   \bottomrule
\end{tabular}
\endgroup

}
\caption{Studies from the Experimental Economics Replication Project.
Shown are the original and replication effect estimates ($\hat\theta_i$)
and $p$-values ($p_i$), the power of the replication study with the 
actual relative sample size $c$, the corresponding relative sample size
$c^{\star}$ calculated 
with the sceptical $p$-value method, $p_{\max}$ and $p_S^*$.}
\label{tbl:EE_studies}
\end{table}




  The studies of \citet{Ericson2011} and \citet{AmbrusGreiner2012}
  deserve closer scrutiny.  While
  $p_{\max} = 0.027$ is the same
  for both studies, {the values of $p_S^*$} differ. In \citet{Ericson2011},
  $p_o = 0.015$ and $p_r =
  0.027$ are relatively close to each other and the red 
  line in Figure~\ref{fig:versus} (left panel) at $c=1/2.4$
   (because we have to reverse the role of
  $p_o$ and $p_r$)
  explains why $p_S^* = 0.032 > p_{\max}$.
  In contrast, there is a larger difference in $p$-values
  ($p_o = 0.027$, $p_r =
  0.006$) in \citet{AmbrusGreiner2012}, and thus
  $p_S^* < p_{\max}$, see the green line in Figure~\ref{fig:versus} (left) at
  $c=3.2$. 
  For this combination of $p$-values, the sceptical $p$-value 
  is smaller than $p_{\max}$ for every value of the relative 
  sample size $c$.
  Of note, in \citet{AmbrusGreiner2012} the two-trials rule
  fails because of the dichotomization at $\alpha = 0.025$,
  but replication success with the sceptical $p$-value is
  achieved ($p_S^* = 0.024$).

\subsection{Sample size calculation}\label{sec:design_eerp}

In the EERP, the replication sample sizes were calculated to reach 
``at least 90\%  power [...] to detect the
original effect size at the [two-sided] 5\% significance level'' 
\citep[p.1434]{Camerer2016}.
We recomputed the exact power values to
 then calculate the required relative sample size $c^{\star}$
with the sceptical $p$-value (see~Table~\ref{tbl:EE_studies}).
The sceptical $p$-value would require
a smaller sample size than the two-trials rule for
12 out of 18 studies in this dataset. Two original studies were non-significant 
at the $\alpha = 0.025$ level \citep{AmbrusGreiner2012, 
  Kuziemko2014}. Sample size calculation based on the sceptical $p$-value
is also
possible for those two studies, where we obtain $c^{\star}>c$.
Note that for these two studies, the sample size $c$ is calculated to achieve significance of the replication study,  
but the two-trials rule will fail anyway.

\section{Discussion}\label{sec:discussion}

We have described a novel statistical framework for the assessment of
replicability, 
stemming from a recently proposed reverse-Bayes approach to assess
replication success \citep{held2020}.
The resulting controlled sceptical $p$-value $p_S^*$ has exact
overall T1E rate of $\alpha^2$ and additionally 
ensures that the conditional and partial T1E rates 
are sufficiently bounded.
%
%
%
%
The two-trials rule can be seen as a special case of the formulation for
$c \downarrow 0$, where $p_S^*$ converges to the maximum of the two
study-specific $p$-values. 
The success region of the sceptical $p$-value
is smooth, shifting
gradually away from the squared one of the two-trials rule for increasing $c$,
thus avoiding the ``double dichotomization'' and offering larger
project power. 
Used in the design of the replication study, the new approach 
requires a smaller sample size than the two-trials rule for already convincing 
original studies. In contrast to the golden version, 
the controlled sceptical $p$-value allows sample size calculation 
for borderline significant and even non-significant original studies.

As the $p$-value $p = (p_S^*)^2$ is a proper $p$-value 
with exact linear T1E control under the intersection null hypothesis, 
a $p$-value function \citep{Fraser2019} could be computed and 
a ``sceptical'' confidence interval could be calculated. 
This would address an
important point raised by \citet{Diggle2020} about the need to accompany the
sceptical $p$-value with suitable estimation procedures to assess the relevance of the
observed effects. 
We plan to consider this in future work.

However, exact overall T1E control comes
at a certain price: the explicit penalization of small relative effect
sizes in the  nominal or golden
versions of the sceptical $p$-value is lost and replication success
may occur even for large shrinkage 
of the replication effect estimate, if
the relative sample size $c$ is large enough.
Our conclusion is that
exact overall T1E control and penalization of small effect sizes are two competing
goals that cannot be achieved by a single criterion. It would therefore be
interesting to extend the recently proposed dual-criterion for
replication studies \citep{Rosenkrantz2021}, 
which simultaneously requires significance and relevance, 
to the sceptical $p$-value.

\hlrev{This paper considers the situation where each original study only has one
replication. Further work will extend the proposed methodology to 
the analysis of multiple replications per original study. A 
natural approach is to perform a meta-analysis of the replication studies and 
use the resulting combined effect estimate (possibly 
allowing for heterogeneity)
as the replication effect estimate.}

\paragraph*{Data and Software Availability} 
Software and data are available in the R-package \texttt{ReplicationSuccess}
available from \texttt{CRAN}. The data is
originally from \url{https://osf.io/pnwuz/}, see \citet[supplement S1]{pawel2020}
for details on data preprocessing. The code to reproduce the analysis and 
figures is available at
\url{https://gitlab.uzh.ch/charlotte.micheloud/framework-for-replicability}.

\paragraph*{Author Contributions} 
CM contributed substantially to research, analysis, coding, and writing. 
FB derived the required null distribution of the sceptical $p$-value,
added further proofs and contributed to writing.
LH designed, performed and supervised research and analysis, wrote parts 
of the code and drafts of the paper.

\paragraph*{Acknowledgments}

LH thanks the University of Zurich for granting a sabbatical leave
that made this research possible.  CM and LH acknowledge
support by the Swiss National Science Foundation (Project \#
189295). We appreciate helpful comments by Rachel Heyard and Samuel Pawel.

\singlespacing
\bibliographystyle{apalike}
\bibliography{antritt}

\begin{appendix}

\counterwithin{figure}{section}

\title{Supporting materials for \\
\hlrev{Assessing replicability with the sceptical $p$-value:} 
Type-I error control and sample size planning} 
\ifcase\blinded 
\author{} \or 
\author{Charlotte Micheloud, Fadoua Balabdaoui and Leonhard Held}
\fi

\maketitle

\doublespacing

  \section{The distribution of $Y= \min\{z_o^2, z_r^2\}$}\label{app:distYmin}
Under the intersection null $X_1=z_o^2$ and $X_2=z_r^2$ are independent $\chi^2(1)$
random variables with cdf $ F_X(x) = 2 \, \Phi(\sqrt{x}) - 1$.
The cdf of the minimum $Y = \min\{X_1, X_2\}$ of two iid
random variables $X_1$ and $X_2$ with cdf $F_X(x)$ has cdf
$F_Y(y) = 1-[1-F_X(y)]^2 = 1-[1-2 \, \Phi(\sqrt{y}) + 1]^2 = 1-4 \, [1-\Phi(\sqrt{y})]^2$, so we obtain Equation~\eqref{eq:mincdf} from the main 
manuscript.

\section{The null distribution of $Y=z_S^2$}\label{app:distY}
Throughout we assume that $z_o$ and $z_r$ are independent standard normal variables, so 
$X_1=z_o^2$ and $X_2=z_r^2$  are i.i.d.~$\sim \chi^2{(1)} \stackrel{d}{=} \Ga(1/2, 1/2)$. 
\subsection{The case  $c < 1$}
Consider $Y=z_S^2$ to be
  equal to the smallest positive root of the equation 
\begin{equation}\label{eq:eq1b}
  \left(\frac{X_1}{Y}-1\right) 
  \left(\frac{X_2}{Y}-1\right) = c .
\end{equation}
    After some algebra, we find for $c<1$ the solution
\begin{eqnarray*}
Y &=&  \frac{1}{2(1-c)}  \left(  X_1 + X_2  - \sqrt{(X_1 + X_2)^2  -  4 (1-c)  X_1 X_2}  \right) \\
& =   &  \frac{ X_1 + X_2}{2(1-c)}  \left( 1  - \sqrt{1-  4 (1-c)  \frac{X_1 X_2}{(X_1  + X_2)^2}}  \right)  \\
 & = &   \frac{S}{2(1-c)} \left( 1  - \sqrt{1-  4 (1-c)  R (1- R)}  \right) \\
 & = &   \frac{S}{2(1-c)} \left( 1  - \sqrt{1-  (1-c)  B}  \right)
 \end{eqnarray*}
where 
\begin{eqnarray*}
S = X_1 +  X_2, \quad R =  \frac{X_1}{X_1 + X_2} \   \  \textrm{and}  \  \  B=4R(1-R).
\end{eqnarray*}
A well-known result  \citep[Section 4.7, Exercise 14]{GrimmettStirzaker2001} says that if $X_1$ and $X_2$ are independent such that $ X_i \sim \Ga(\alpha_i, \beta)$, $i=1,2$, then 
\begin{eqnarray*}
S = X_1 + X_2  \sim \Ga(\alpha_1 +\alpha_2, \beta)   \ \  \perp \!\!\! \perp \  \    R =  \frac{X_1}{X_1+X_2}  \sim \Be(\alpha_1, \alpha_2).
\end{eqnarray*} 
We have $X_i \sim \Ga(1/2, 1/2)$ and therefore
$S \sim \Ga(1, 1/2) \stackrel{d}{=} \Exp(1/2)$ and
$R \sim \Be(1/2,1/2)$ are independent.  The $\Be(1/2,1/2)$ is also
known as the 
$\ArcS(0,1)$ distribution on the support $[0,1]$ \citep{Rogozin2001}.
The general arcsine distribution $\ArcS(a,a+b)$ on the support $[a,a+b]$, $a \in \R$ and $b>0$, is
obtained by the linear transformation $a + b X$ with $X \sim \ArcS(0,1)$. If $b < 0$, 
then $a + b X \sim \ArcS(a+b, a)$. 
The arcsine distribution has the property 
\[
  X \sim \ArcS(-1,1) \quad \Rightarrow \quad X^2 \sim \ArcS(0,1) 
\]
which implies that
$B=4R(1-R) = 1- (1-2R)^2 \sim \Be(1/2,1/2)$ holds with $S$ and $B$ independent.

It is clear that $F_c(y) = 0$ for $y \le 0$. For $y > 0$,  we now obtain
\begin{eqnarray}
F_c(y)  & = & \P(Y \le y)  \nonumber \\
        & = &   \P\left(  S \left( 1  - \sqrt{1- (1-c)  B}  \right)   \le 2(1-c)   y \right)  \nonumber \\
& =  & \int_0^1  \P\left(  S  \le \frac{2(1-c) y}{1  - \sqrt{1- (1-c) t} }  \right)  \frac{\Gamma(1)}{\Gamma(1/2)^2}  t^{-1/2}  (1-t)^{-1/2} dt \nonumber \\
& = &  1  -  \frac{1}{\pi} \int_0^1 \exp\left (- \frac{(1-c) y}{1  - \sqrt{1+ (1-c)  t} }   \right)\frac{1}{\sqrt{t(1-t)}}   dt \label{eq:cdf1}
\end{eqnarray}
using independence of $S$ and $B$ and 
the fact that the cdf of $\Exp(1/2)$ is given by $t \mapsto 1-\exp(-t/2), t > 0$ and that $1/\pi \int_0^1 1/\sqrt{r(r-1)}  dr =1$ (the density of  $\Be(1/2,1/2)$ integrates to $1$).   
  \subsection{The case $c > 1$}

The solution of equation \eqref{eq:eq1b} is in this case
\begin{eqnarray*}
Y 
& = & \frac{S}{2 (c-1)}  \left(  \sqrt{1 +  (c-1) B} - 1  \right)
\end{eqnarray*}
with $S$ and $B$ defined as before. 
Thus,  for $  y > 0$,
\begin{eqnarray}
F_c(y) & = &  1  -  \frac{1}{\pi} \int_0^1 \exp\left (- \frac{(c-1) y}{\sqrt{1+ (c-1)  t} -1}   \right)\frac{1}{\sqrt{t(1-t)}}   dt. \label{eq:cdf2}
\end{eqnarray}
\\[.5cm]

\subsection{The expectation of $Y$}\label{app:expY}
For $c=1$ we obtain $\E(Y)=0.25$ from
$Y \sim \Ga(1/2, 2)$.

For $c=0$ we have 
\begin{eqnarray*}
\E(Y) & = & \min\{X_1, X_2\} \\
& = & \frac{1}{2}(X_1 + X_2 - \abs{X_1 - X_2}) \\
& = & \frac{1}{2}(z_o^2 + z_r^2 - \abs{z_o^2 - z_r^2}) \\
& = & \frac{1}{2}(z_o^2 + z_r^2 - \abs{z_o - z_r} \abs{z_o + z_r}). 
\end{eqnarray*}
Now $\Cov(z_o - z_r, z_o + z_r) = \Var(z_o) - \Var(z_r) = 0$ and hence
${z_o - z_r}$ and ${z_o + z_r}$ are independent $\Nor(0,2)$ variables. 
This implies that $\abs{z_o - z_r}$ and $\abs{z_o + z_r}$ are also independent
and identically distributed according to a 
half-normal distribution with expectation $2/\sqrt{\pi}$. Thus, 
\begin{eqnarray*}
\E(Y) & = & \frac{1}{2}(\E(z_o^2) + \E(z_r^2) - \E(\abs{z_o - z_r}) \E(\abs{z_o + z_r})) \\
& = & \frac{1}{2}(1 + 1 - (2/\sqrt{\pi})^2) \\
& = & 1-2/\pi \approx 0.36. 
\end{eqnarray*}

For $0 < c<1$, the random variable $Y$ has cdf \eqref{eq:cdf1} with expectation
\begin{eqnarray*}
E(Y) & = & \int_0^\infty (1-F_c(y)) dy \\
  & = & \frac{1}{\pi} \int_0^\infty \int_0^{1} \exp \left(-\frac{(1-c)y}{1-\sqrt{1-(1-c)t}} \right) \frac{1}{\sqrt{t(1-t)}} dt \, dy \\
   & = & \frac{1}{\pi} \int_0^{1} \frac{1}{\sqrt{t(1-t)}} \int_0^\infty \exp \left(-\frac{(1-c)y}{1-\sqrt{1-(1-c)t}} \right)  dy \, dt .
\end{eqnarray*}
Now
\begin{eqnarray*}
 \int_0^\infty \exp \left(-\frac{(1-c)y}{1-\sqrt{1-(1-c)t}} \right)  dy 
& = &         \frac{1-\sqrt{1-(1-c)t}}{1-c} 
\end{eqnarray*}
and therefore
\begin{eqnarray*}
E(Y) & = & 
\frac{1}{\pi(1-c)} \int_0^{1} \frac{1-\sqrt{1-(1-c)t}}{\sqrt{t(1-t)}}   dt .
\end{eqnarray*}
For $c>1$ we obtain with \eqref{eq:cdf2}
\begin{eqnarray*}
E(Y)    & = & \frac{1}{\pi(c-1)} \int_0^{1} \frac{\sqrt{1-(c-1)t}-1}{\sqrt{t(1-t)}}   dt .
\end{eqnarray*}

\section{The limit of the $p$-value as $c \to \infty$}\label{app:plimit}

Recall that for $c>1$
\[
Z =  \frac{S}{2 (c-1)}  \left(   -1  + \sqrt{1  +   (c-1) B}  \right)
\]
with $S = X_1 + X_2$ and $B =  4 R(1-R)$.  The \hl{two-sided} $p$-value 
$4p(c) = 1-F_c(z_S^2)$
can be factorized as 
\begin{eqnarray*}
 4p(c) 
                   & = &  \frac{1}{\pi} \int_0^{1}\exp\left (- \frac{(c-1) Z}{\sqrt{1+  (c-1) t}   -1}   \right) \frac{1}{\sqrt{t(1-t)} }  dt  \\
                   & =&   \frac{1}{\pi} \int_0^{\eta}\exp\left (- \frac{(c-1) Z}{\sqrt{1+  (c-1) t}   -1}   \right) \frac{1}{\sqrt{t(1-t)} }  dt  \\
&+ &  \frac{1}{\pi} \int_{1-\eta}^{1}\exp\left (- \frac{(c-1) Z}{\sqrt{1+  (c-1) t}   -1}   \right) \frac{1}{\sqrt{t(1-t)} }  dt  \\
&+ &    \frac{1}{\pi} \int_{\eta}^{1-\eta}\exp\left (- \frac{(c-1) Z}{\sqrt{1+  (c-1) t}   -1}   \right) \frac{1}{\sqrt{t(1-t)} }  dt  \\
& = &   A(\eta,c) +  B(\eta, c) +  C(\eta, c).
\end{eqnarray*}
It is  easy to see that
\begin{eqnarray*}
\sup_{c > 1} A(\eta,c)   \le  \int_0^{\eta} \frac{1}{\sqrt{t(1-t)} }  dt   \searrow 0, \  \ \textrm{as $\eta \searrow 0$}.
\end{eqnarray*}
Similarly,
\begin{eqnarray*}
 \sup_{c > 1}   B(\eta,c)   \le  \int_{1-\eta}^1 \frac{1}{\sqrt{t(1-t)} }  dt   \searrow 0, \  \ \textrm{as $\eta \nearrow 1$}.
\end{eqnarray*}
Now, we want to show that
\begin{eqnarray*}
 \lim_{c \to \infty}  C(\eta, c)  =   \frac{1}{\pi}   \int_{\eta}^{1-\eta}\exp\left (- \frac{S \sqrt{ B}}{2\sqrt t}\right) \frac{1}{\sqrt{t(1-t)} }  dt.
\end{eqnarray*}
Let us put
\begin{eqnarray*}
\Delta(\eta, c)=  \frac{1}{\pi}  \int_{\eta}^{1-\eta}  \left \{  \exp\left(\frac{ S \left(-1  + \sqrt{1+ (c-1) B } \right)}{2 \left(\sqrt{1+  (c-1) t}   -1\right)}   \right)   - \exp\left(\frac{ S \sqrt B}{2\sqrt t} \right) \right \}  \frac{1}{\sqrt{t(1-t)} }  dt.
\end{eqnarray*}
We show now that $\lim\limits_{c  \to \infty} \Delta(\eta, c) =0$.   It is enough to show that for a $\eta > 0$ small enough,  
\begin{eqnarray*}
\lim_{c \to \infty} \sup_{t  \in [\eta, 1- \eta] }  \abs{ \, \exp\left(\frac{ S \left(-1  + \sqrt{1+ (c-1) B } \right)}{2 \left(\sqrt{1+  (c-1) t}   -1\right)}   \right)   - \exp\left(\frac{ S \sqrt B}{2\sqrt t} \right)  } = 0.
\end{eqnarray*}
By the mean-value theorem, $ \exp(y) - \exp(x)  = \exp(\theta^*_{x, y}) (y - x) $ for some $\theta^*_{x, y}$ between $x$ and $y$.   Thus, $\abs{  \exp(y) - \exp(x) } \le \exp(\max(x, y))  \abs{ y - x }$.   In what follows, 
\begin{eqnarray*}
x =  \frac{S \sqrt{B}}{2 \sqrt{t}} \ \ \ \textrm{and} \  \  \  y =  \frac{S(-1  + \sqrt{1+(c-1) B})}{2 \left(\sqrt{1+(c-1)t } -1\right)}.
\end{eqnarray*}
We have 
\begin{eqnarray*}
y & = &  \frac{S}{2}   \times \frac{\sqrt{B(c-1)}  \left( \sqrt{1 + \frac{1}{(c-1)B}}  - \frac{1}{\sqrt{(c-1)B}}  \right)}{\sqrt{(c-1) t}  \left( \sqrt{1+ \frac{1}{\sqrt{(c-1)t}}}  -  \frac{1}{\sqrt{(c-1) t}}  \right) }  \\
& = &  \frac{S}{2 \sqrt t}   \times  \frac{\sqrt{1 + \frac{1}{(c-1)B}}  - \frac{1}{\sqrt{(c-1)B}}}{ \sqrt{1 + \frac{1}{(c-1)t}}   -  \frac{1}{\sqrt{(c-1) t}}  }  \\
& = &  x    \times  \frac{\sqrt{1 + \frac{1}{(c-1)B}}  - \frac{1}{\sqrt{(c-1)B}}}{ \sqrt{1 + \frac{1}{(c-1)t}}   -  \frac{1}{\sqrt{(c-1) t}}  }
\end{eqnarray*}
where 
\begin{eqnarray*}
\frac{\sqrt{1 + \frac{1}{(c-1)B}}  - \frac{1}{\sqrt{(c-1)B}}}{ \sqrt{1 + \frac{1}{(c-1)t}}   -  \frac{1}{\sqrt{(c-1) t}}  }   &= &   \frac{\sqrt{1 + \frac{1}{(c-1)t}}  + \frac{1}{\sqrt{(c-1)t}}}{ \sqrt{1 + \frac{1}{(c-1)B}}   +  \frac{1}{\sqrt{(c-1) B}}  }  \\
&  \le  &   \sqrt{1 + \frac{1}{(c-1)t}}  + \frac{1}{\sqrt{(c-1)t}}  \\
& \le &   \sqrt{1 + \frac{1}{(c-1)\eta}}  + \frac{1}{\sqrt{(c-1)\eta}}  \\
& \le & \frac{3}{2}
\end{eqnarray*}
for  $c$ large enough.   It follows that 
\begin{eqnarray}\label{Term}
\abs{ \exp(y)  - \exp(x) }  & \le   &  \exp\left(\frac{3}{2} x\right)  \abs{ y -  x }   \nonumber \\
   & \le  &    \exp\left(\frac{3 S \sqrt{B}}{4\sqrt{\eta}}\right)   \abs{  x \frac{\sqrt{1 + \frac{1}{(c-1)B}}  - \frac{1}{\sqrt{(c-1)B}}}{ \sqrt{1 + \frac{1}{(c-1)t}}   -  \frac{1}{\sqrt{(c-1) t}}  }    -   x   } \nonumber \\ 
& \le &  \exp\left(\frac{3 S \sqrt{B}}{4\sqrt{\eta}}\right) \frac{S \sqrt{B}}{2 \sqrt{\eta}}  \abs{   \frac{\sqrt{1 + \frac{1}{(c-1)B}}  - \frac{1}{\sqrt{(c-1)B}}}{ \sqrt{1 + \frac{1}{(c-1)t}}   -  \frac{1}{\sqrt{(c-1) t}}  }    - 1 }
\end{eqnarray}
where 
\begin{eqnarray*}
 \frac{\sqrt{1 + \frac{1}{(c-1)B}}  - \frac{1}{\sqrt{(c-1)B}}}{ \sqrt{1 + \frac{1}{(c-1)t}}   -  \frac{1}{\sqrt{(c-1) t}}  }    - 1  =   \frac{\sqrt{1 + \frac{1}{(c-1)t}} +  \frac{1}{\sqrt{(c-1) t}}}{\sqrt{1 + \frac{1}{(c-1)B}}  + \frac{1}{\sqrt{(c-1)B}}}    - 1
\end{eqnarray*}
implying that 
\begin{eqnarray*}
\frac{\sqrt{1 + \frac{1}{(c-1)(1-\eta)}} +  \frac{1}{\sqrt{(c-1) (1-\eta)}}}{\sqrt{1 + \frac{1}{(c-1)B}}  + \frac{1}{\sqrt{(c-1)B}}}    - 1  \le \frac{\sqrt{1 + \frac{1}{(c-1)B}}  - \frac{1}{\sqrt{(c-1)B}}}{ \sqrt{1 + \frac{1}{(c-1)t}}   -  \frac{1}{\sqrt{(c-1) t}}  }    - 1  \le \frac{\sqrt{1 + \frac{1}{(c-1)\eta}} +  \frac{1}{\sqrt{(c-1) \eta}}}{\sqrt{1 + \frac{1}{(c-1)B}}  + \frac{1}{\sqrt{(c-1)B}}}    - 1.
\end{eqnarray*}
As the terms in both sides of the inequality converge to $0$ as  $c \nearrow \infty$ we conclude that the absolute value of the term in the middle converges to $0$ and hence the right side of the inequality in (\ref{Term}) has also to converge to $0$ where the convergence is uniform for $t \in [\eta, 1-\eta]$.
It follows that 
$\lim\limits_{c \nearrow \infty}  \Delta(\eta, c)  =0$ 
and hence,  for $\eta$ small enough
\begin{eqnarray*}
\lim_{c \nearrow \infty}  C(\eta, c)  =  \frac{1}{\pi}   \int_{\eta}^{1-\eta}\exp\left (- \frac{S \sqrt{ B}}{2\sqrt t}\right) \frac{1}{\sqrt{t(1-t)} }  dt.
\end{eqnarray*}
Therefore,  and since $4p(c) $ does not depend on $\eta$
\begin{eqnarray}
\lim_{c \nearrow  \infty} 4p(c)  & = &   \lim_{\eta \searrow 0}  \lim_{c \nearrow  \infty} 4p(c)  \nonumber \\
& = & 0 +  0  + \frac{1}{\pi}   \int_{0}^{1}\exp\left (- \frac{S \sqrt{ B}}{2\sqrt t}\right) \frac{1}{\sqrt{t(1-t)} }  dt \nonumber \\
& = & \frac{1}{\pi}   \int_{0}^{1}\exp\left (- \frac{S \sqrt{ B}}{2\sqrt t}\right) \frac{1}{\sqrt{t(1-t)} }  dt \label{eq:mynumber}
\end{eqnarray}
which finishes the proof because
\begin{eqnarray*}
W= \frac{S \sqrt{B}}{2}  =  \sqrt{X_1  X_2}   =  \abs{ z_o z_r}.
\end{eqnarray*}
\hfill $\Box$

\section{Uniform distribution of the limiting $p$-value}\label{app:uniform}

We want to show that the $p$-value from \eqref{eq:mynumber} is uniformly distributed on $[0,1]$. To this aim, we need to determine  the density of $S \sqrt B/2 = \abs{ z_o z_r}$
with $z_o, z_r$ i.i.d $\sim \mathcal{N}(0,1)$.   
The density of the variable $V = z_o z_r$ is
\begin{eqnarray*}
f_V(v)  = \frac{1}{\pi}  K_0(\abs{ v })
\end{eqnarray*}
with $K_0$ is second class zero order modified Bessel function \citep{Craig1936}.  
Now, the cumulative distribution of $W$ is given for $w > 0$  by 
$F_W(w)  =  2 F_V(w) - 1$,
implying that density of $W$ is given by
\begin{eqnarray*}
f_W(w)  =  2 f_V(w)    =     \frac{2}{\pi}  K_0(w), \ \   \textrm{for $w > 0$}.
\end{eqnarray*}
To show that the $p$-value from \eqref{eq:mynumber} is uniformly distributed on $[0,1]$ it is enough to show that 
1 minus the $p$-value from \eqref{eq:mynumber} is uniformly distributed on $[0,1]$. This is equivalent to showing that the density of $W$ is also equal to   the derivative of 
\begin{eqnarray*}
w \mapsto 1-\frac{1}{\pi}   \int_{0}^{1}\exp\left (- \frac{w}{\sqrt t}\right) \frac{1}{\sqrt{t(1-t)} }  dt.
\end{eqnarray*}
Since this derivative is 
\begin{eqnarray*}
w \mapsto \frac{1}{\pi}   \int_{0}^{1}\exp\left (- \frac{w}{\sqrt t}\right) \frac{1}{t\sqrt{1-t} }  dt,
\end{eqnarray*}
we need to show that 
\begin{eqnarray*}
\frac{1}{\pi}   \int_{0}^{1}\exp\left (- \frac{w}{\sqrt t}\right) \frac{1}{t\sqrt{1-t} }  dt   =  \frac{2}{\pi}  K_0(w)
\end{eqnarray*}
for all $w > 0$ or equivalently
\begin{eqnarray}\label{Id}
\int_{0}^{1}\exp\left (- \frac{w}{\sqrt t}\right) \frac{1}{t\sqrt{1-t} }  dt   =  2 K_0(w).
\end{eqnarray}
It is known that for $n > -1/2$, the second class modified Bessel function of order $n$ is for $x > 0$ 
\begin{eqnarray*}
K_n(x) =  \frac{\sqrt{\pi}}{\Gamma(n + 1/2)}   \left(\frac{1}{2}  z\right)^n  \int_1^\infty \exp(-xt)(t^2-1)^{n-1/2}  dt.
\end{eqnarray*}
 Thus, for $n=0$,
\begin{eqnarray*}
K_0(x) =  \int_1^\infty \frac{\exp(-xt)}{\sqrt{t^2-1}}  dt.
\end{eqnarray*}
Thus, to show the identity in (\ref{Id}), it is enough to show that 
\begin{eqnarray*}
\int_{0}^{1}\exp\left (- \frac{w}{\sqrt t}\right) \frac{1}{t\sqrt{1-t} }  dt   =  2  \int_1^\infty \frac{\exp(-wt)}{\sqrt{t^2-1}}  dt
\end{eqnarray*}
for all $w > 0$.  Using the change of variable $u = 1/\sqrt t$ we obtain
\begin{eqnarray*}
\int_{0}^{1}\exp\left (- \frac{w}{\sqrt t}\right) \frac{1}{t\sqrt{1-t} }  dt   & =   &   \int_1^\infty \exp(-wu) u^2  \frac{1}{\sqrt{1-1/u^2}}  \frac{2}{u^3} du  \\
& = &  2 \int_1^\infty \frac{\exp(-wu)}{\sqrt{u^2 - 1}} du
\end{eqnarray*}
which completes the proof. \hfill $\Box$

\section{Derivative of the sceptical $p$-value}\label{sec:deriv}
\hl{
In the following, we write $u= c-1$. Then, we have 
\begin{eqnarray*}
p_S^*  &=  & p_S^*(u)=  \frac{1}{4\pi}  \int^1_0  \exp \left  (- z^2_A   \frac{\sqrt{1+u  z_H^2/z^2_A}  -1}{\sqrt{1+ u t}-1}  \right)   \frac{1}{\sqrt{t(1-t)}} dt  \\\
& = &  \frac{1}{4\pi}  \int^1_0  \exp \left  (- z^2_A   \frac{\sqrt{1+u  B}  -1}{\sqrt{1+ u t}-1}  \right)   \frac{1}{\sqrt{t(1-t)}} dt
\end{eqnarray*}
with  $ B =  z_H^2/z^2_A$.  It is not difficult to show that the function on the left can be differentiated with respect to $u$ under the sign integral.  Then,  for $u \in \mathbb R$
\begin{eqnarray*}
 p_S^{'*}(u) =  -\frac{z^2_A}{4\pi}  \int^1_0  \exp \left  (- z^2_A  \frac{\sqrt{1+u  B}  -1}{\sqrt{1+ u t}-1}  \right)   \frac{d}{du}  \left\{ \frac{\sqrt{1+u  B}  -1}{\sqrt{1+ u t}-1}\right \}   \frac{1}{\sqrt{t(1-t)}} dt.
\end{eqnarray*}
We  compute
\begin{eqnarray*}
\frac{d}{du}  \left\{ \frac{\sqrt{1+u  B}  -1}{\sqrt{1+ u t}-1}\right \} &  =  &  \frac{B(1+ut - \sqrt{1+ut}) - t (1+u B - \sqrt{1+u B})}{2 (\sqrt{1+ut}-1)^2 \sqrt{1+uB} \sqrt{1+ut}}  \\
& = &  \frac{ B- t - B \sqrt{1+ut}  +  t \sqrt{1+u B}}{2 (\sqrt{1+ut}-1)^2 \sqrt{1+uB} \sqrt{1+ut}} \\
& = &  \frac{ \left(B- t - B \sqrt{1+ut}  +  t \sqrt{1+u B}  \right) (\sqrt{1+ut}+1)^2  }{2 u^2 t^2 \sqrt{1+uB} \sqrt{1+ut}}. 
\end{eqnarray*}
Thus, the derivative of the sceptical $p$-value for $c > 0$ and $c \ne 1$ is given by
\begin{eqnarray*}\label{derivc}
-\frac{z^2_A}{8\pi} \int_0^1   \exp \left  (- z^2_A  \frac{\sqrt{1+(c-1)  B}  -1}{\sqrt{1+ (c-1) t}-1}  \right)  k_{c, B}(t) dt
\end{eqnarray*}
with
\begin{eqnarray*}\label{k}
 k_{c, B}(t) =  \frac{ \left(B- t - B \sqrt{1+(c-1)t}  +  t \sqrt{1+(c-1) B}  \right) (\sqrt{1+(c-1)t}+1)^2  }{ (c-1)^2 \sqrt{1+(c-1)B} \sqrt{1+(c-1)t}}   \frac{1}{t^{5/2} \sqrt{1-t}} dt.
\end{eqnarray*}
For $c=1$, the derivative is given by
\begin{eqnarray*}\label{deriv1}
-\frac{z^2_A}{16\pi} \int_0^1   \exp \left  (- z^2_A \frac{B}{t}  \right)   \frac{t-B}{t^{3/2} \sqrt{1-t}} dt.
\end{eqnarray*}
}

\section{Monotonicity of the sceptical $p$-value for the case $z_o=z_r$}\label{sec:monotone}

\hl{
In the case $z_o = z_r$, we  have that
\begin{eqnarray*}
z^2_S  = \frac{z^2_o}{c-1}  (\sqrt{c}  - 1)
\end{eqnarray*}
and hence the one-sided $p$-value $p$ is
\begin{eqnarray*}
p= \left[1- F_c(z^2_S)\right]/4   &  =  &   \frac{1}{4\pi}  \int_0^1  \exp\left(-z^2_o \frac{\sqrt{c} -1}{\sqrt{1 + (c-1)t} -1 }   \right)   
\frac{1}{\sqrt{t(1-t)}}  dt \\
& =  &    \frac{1}{4\pi}  \int_0^1  \exp\left(-z^2_o  \frac{( \sqrt{c} -1) \sqrt{1 + (c-1)t} +1 }{(c-1) t }   \right)   
\frac{1}{\sqrt{t(1-t)}}  dt  \\
& = &   \frac{1}{4\pi}  \int_0^1  \exp\left(-\frac{z^2_o}{t}  \frac{ \sqrt{1 + (c-1)t} +1 }{\sqrt{c} +1}   \right)   
\frac{1}{\sqrt{t(1-t)}}  dt  \\
& = &  \frac{1}{4\pi}  \int_0^1  \exp\left(-\frac{z^2_o}{t} \psi_t(c)   \right)   
\frac{1}{\sqrt{t(1-t)}}  dt
\end{eqnarray*}
with
\begin{eqnarray*}
\psi_t(c) = \frac{ \sqrt{1 + (c-1)t} +1 }{\sqrt{c} +1}, (t, c)  \in [0,1] \times [0, \infty).
\end{eqnarray*}
If we show that $\psi_t$is monotone decreasing in $c$ for all $t \in [0,1]$, then this would imply the \hl{sceptical $p$-value $p_S^* = \sqrt{p}$} is monotone increasing in $c$.  Fix $t \in [0,1]$. The derivative of $\psi_t$ for $ c > 0$ is given by 
\begin{eqnarray*}
\psi'_t(c) & =   &   \frac{  \frac{t }{2 \sqrt{1 + (c-1)t}}   (\sqrt{c} +1)   -  \frac{\sqrt{1 + (c-1)t} +1}{2 \sqrt c}}{(\sqrt c + 1)^2}  \\
& =  &   \frac{ t  (\sqrt{c} +1) \sqrt c  -  (\sqrt{1 + (c-1)t} +1)\sqrt{1 + (c-1)t}} {2 \sqrt{1 + (c-1)t} \sqrt c (\sqrt c + 1)^2} \\
& = &   \frac{ t  (c +\sqrt c)   -  (1 + (c-1)t) -\sqrt{1 + (c-1)t}} {2 \sqrt{1 + (c-1)t} \sqrt c (\sqrt c + 1)^2}  \\
& = &  \frac{t (\sqrt c +1)  - 1   - \sqrt{1 + (c-1)t} }{2 \sqrt{1 + (c-1)t} \sqrt c (\sqrt c + 1)^2}.
\end{eqnarray*}
We show now that $ t (\sqrt c +1)    \le \sqrt{1 + (c-1)t}  +1 $ for all $c > 0$.  This is equivalent to showing that $t (\sqrt c +1)   - 1   \le \sqrt{1 + (c-1)t}$. If $t (\sqrt c+1) - 1 \le 0$, then this is obviously true. Suppose now that $t (\sqrt c + 1) - 1 > 0$. Then, it is enough to show that 
\begin{eqnarray*}
\left(t (\sqrt c+1) - 1\right)^2 \le 1 + (c-1)  t. 
\end{eqnarray*}
We compute
\begin{eqnarray*}
\left(t (\sqrt c+1) - 1\right)^2  - (1 + (c-1)  t)   & =   &   t^2 (c + 2\sqrt c+1)   -  2t \sqrt c - 2t    - ct +  t \\ 
& = &   t^2 (c + 2\sqrt c+1)   -  2t \sqrt c    -  ct -  t  \\
& =  &   (t^2 - t)  \left(c + 2 \sqrt c   +1  \right)    =  t (t-1)  (\sqrt c + 1)^2  \le 0
\end{eqnarray*}
and the proof is completed.}

\section{Design of replication studies}\label{sec:predPower}

\begin{figure}[!ht]
\begin{knitrout}
\definecolor{shadecolor}{rgb}{0.969, 0.969, 0.969}\color{fgcolor}
\includegraphics[width=\maxwidth]{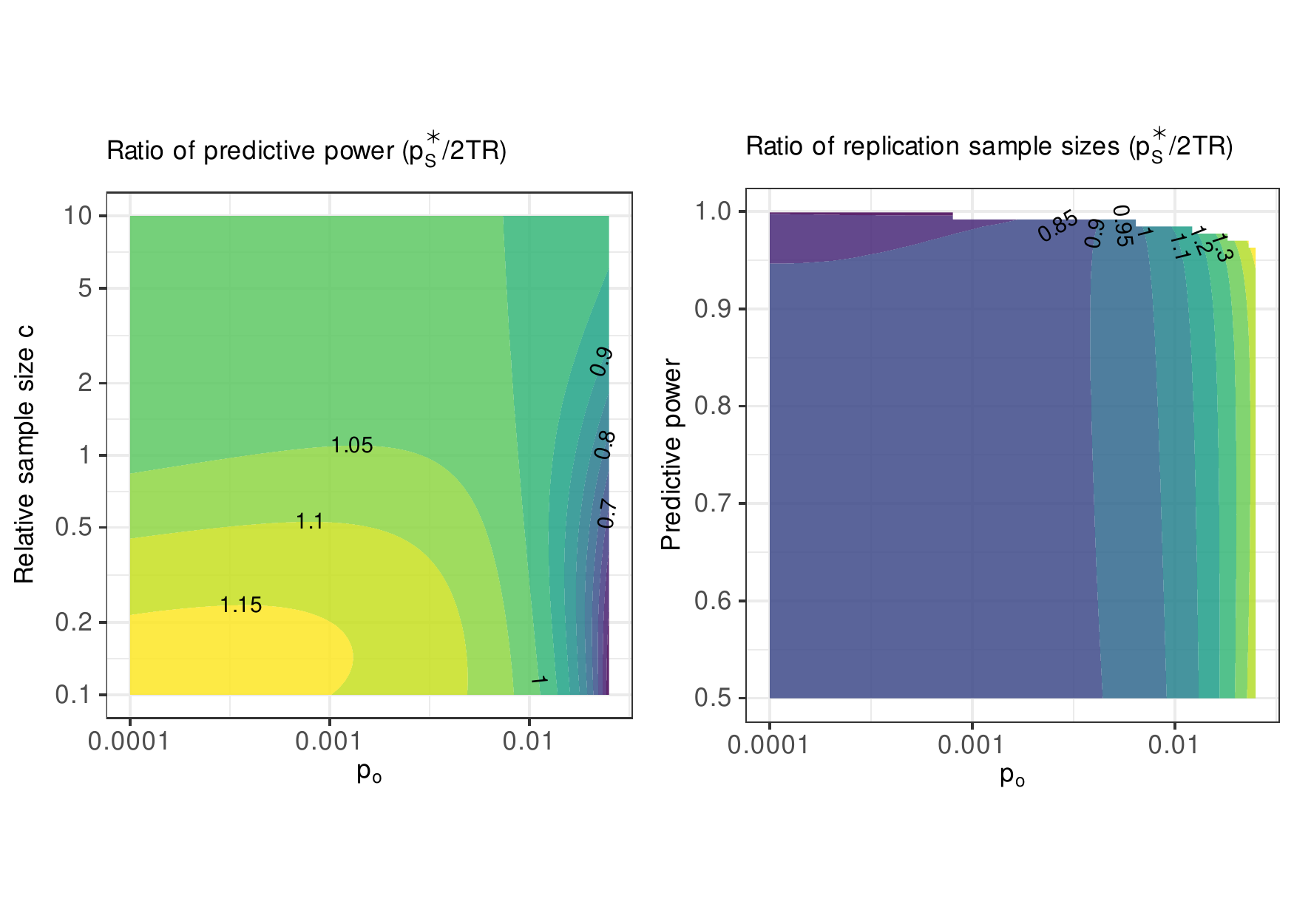} 
\end{knitrout}
\vspace{-1cm}
\caption{
\hl{Ratio of predictive power (left) and replication 
sample sizes (right) between the
   sceptical $p$-value and the two-trials rule as a function 
  of the original $p$-value $p_o$ and the relative sample size 
  $c$ (left) or the predictive power (right)
  for $\alpha = 0.025$.}
  }
\label{fig:powRepPred}
\end{figure}

\begin{figure}[!h]
\begin{knitrout}
\definecolor{shadecolor}{rgb}{0.969, 0.969, 0.969}\color{fgcolor}
\includegraphics[width=\maxwidth]{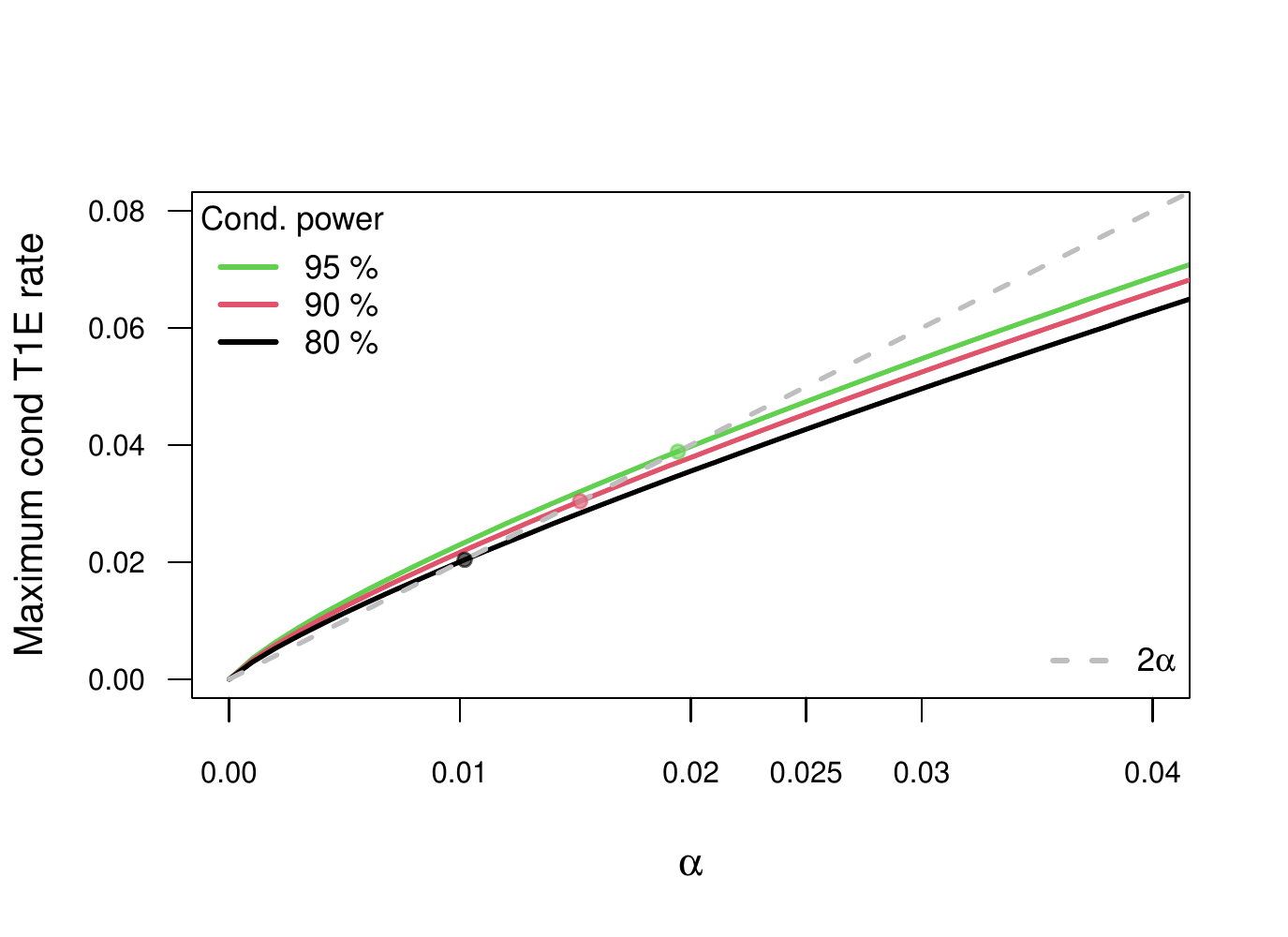} 
\end{knitrout}
\caption{Maximum conditional T1E rate of the controlled sceptical 
$p$-value as a function of the level $\alpha$.}
\label{fig:maxcondt1e}
\end{figure}

\begin{figure}[!h]
  \centering
\begin{knitrout}
\definecolor{shadecolor}{rgb}{0.969, 0.969, 0.969}\color{fgcolor}
\includegraphics[width=\maxwidth]{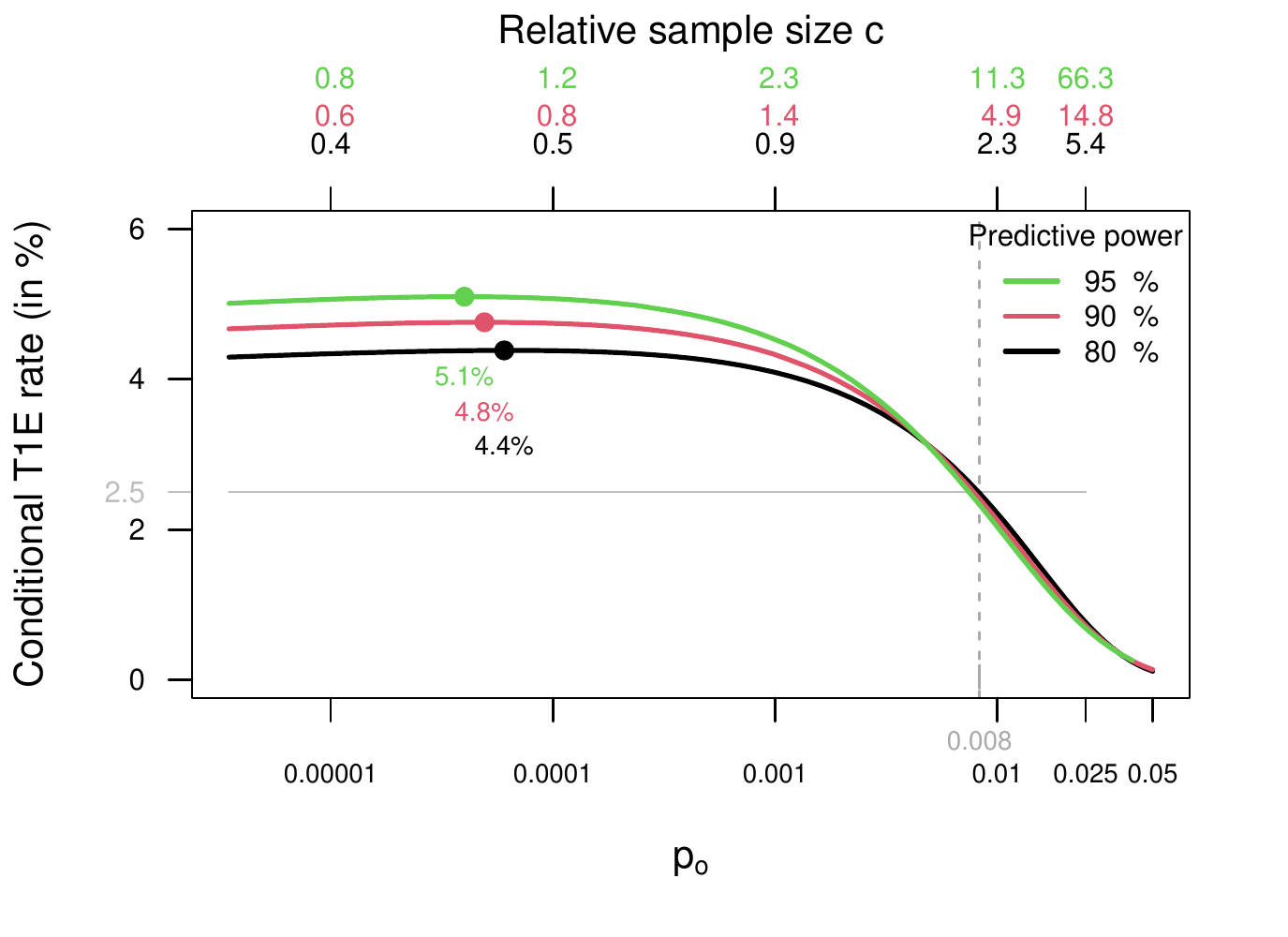} 
\end{knitrout}
\caption{Conditional Type-I error rate as a function of the original $p$-value
$p_o$. 
The replication sample size is calculated with the sceptical $p$-value method 
to reach a predictive power of 80\%, 90\%, and 95\% with $\alpha = 0.025$.
Each dot represents the upper bound for the conditional T1E rate with the 
respective power.
The gray horizontal line indicates the T1E rate of the two-trials rule.}
\label{fig:pred_rep_t1e}
\end{figure}

\begin{figure}[!h]
\centering
\begin{knitrout}
\definecolor{shadecolor}{rgb}{0.969, 0.969, 0.969}\color{fgcolor}
\includegraphics[width=\maxwidth]{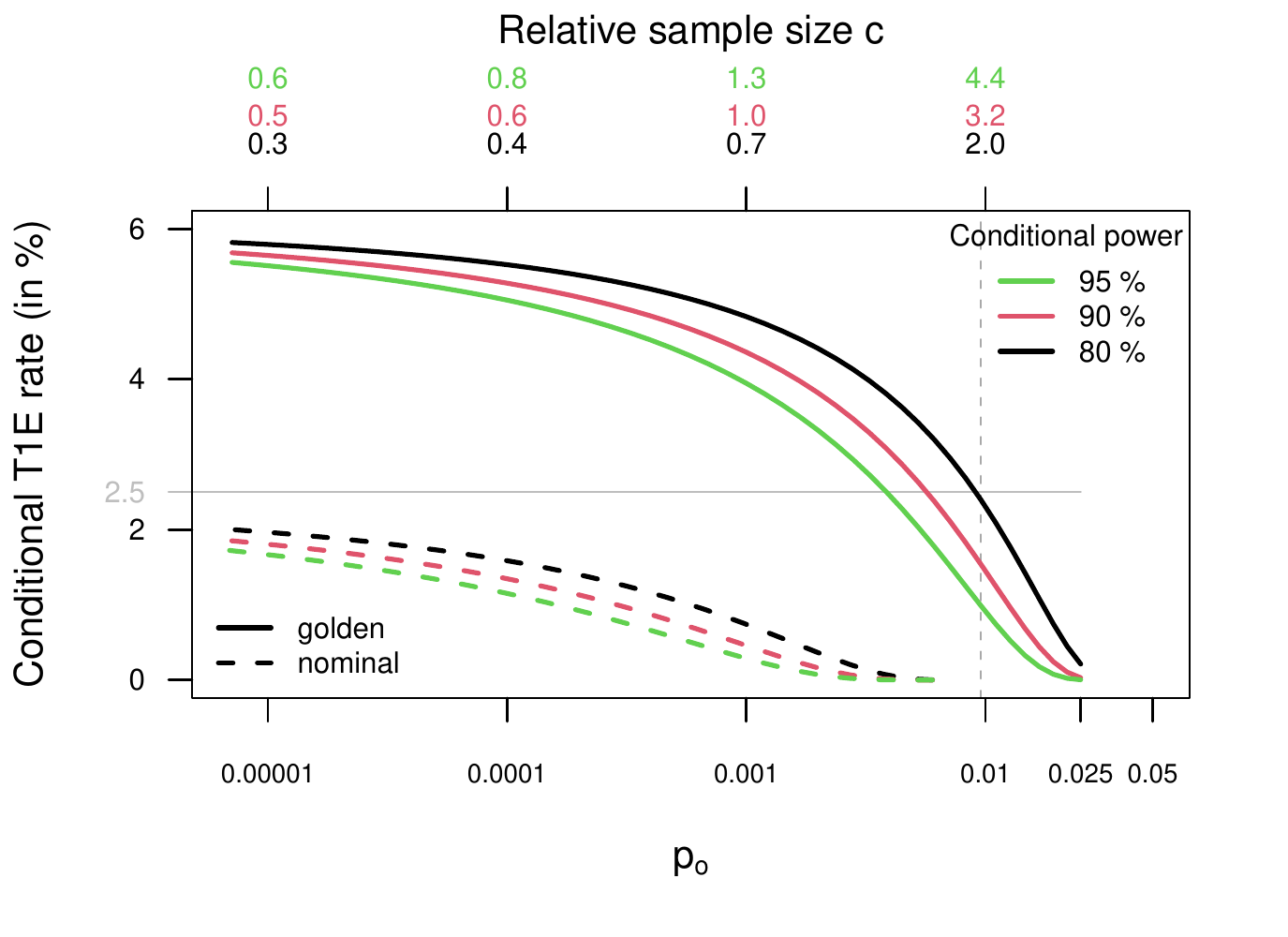} 
\end{knitrout}
\caption{Conditional T1E error rate as a function of the original $p$-value
$p_o$ for the nominal and golden sceptical $p$-values. 
The replication sample size is calculated with the sceptical $p$-value method 
to reach a conditional power of 80\%, 90\%, and 95\% 
with $\alpha = 0.025$. The 
relative sample size $c$ on the top axis is calculated with the 
golden sceptical $p$-value. 
The gray horizontal line indicates the T1E rate of the two-trials rule.}
\label{fig:rep_t1e_nomgold}
\end{figure}

\hl{Figure~\ref{fig:powRepPred} shows the ratio of predictive power and 
replication sample size calculated with the sceptical $p$-value
versus the two-trials rule. Some large predictive power values cannot 
be reached regardless of the sample size \citep{MicheloudHeld2022}. 
This explains the white area (upper-right corner) in the right plot.}
\hlrev{Figure~\ref{fig:maxcondt1e} shows the maximum conditional T1E rate of
the controlled sceptical $p$-value
as a function of $\alpha$ when the sample size is calculated based on a
conditional power of 80\%, 90\% and 95\%, respectively.
The conditional T1E rate is smaller than $2\alpha$ for values of $\alpha$ 
larger than $0.01$, $0.015$ and $0.019$, 
respectively.}
\hl{Figure~\ref{fig:pred_rep_t1e}
shows the conditional T1E rate in the case 
where the replication sample size was calculated based on 
predictive power.
Figure~\ref{fig:rep_t1e_nomgold} shows the conditional T1E rate based on 
nominal and golden sceptical $p$-values.
}

\end{appendix}

\end{document}